\documentclass[12pt]{article}
\usepackage{amssymb,tabularx,multirow,amsmath,natbib, epsfig, threeparttable, amstext, subfigure}

\newcommand{\blind}{1}

\begin{document}

\def\spacingset#1{\renewcommand{\baselinestretch}%
{#1}\small\normalsize} \spacingset{1}


\if1\blind
{
  \title{\bf Basis Function Models for Animal Movement}
  \author{Mevin B. Hooten\thanks{
    \tiny The authors thank Mat Alldredge, Frances Buderman, Marti Garlick, Ephraim Hanks, Bill Link, Brett McClintock, Leslie Mcfarlane, Juan Morales, Jim Powell, Henry Scharf, and Jay Ver Hoef for help for insightful discussions and feedback on the data and research. Funding for this research was provided by NOAA (RWO 103), CPW (TO 1304), and NSF (DMS 1614392).  Any use of trade, firm, or product names is for descriptive purposes only and does not imply endorsement by the U.S. Government.}\hspace{.2cm}\\
    U.S. Geological Survey \\ Colorado Cooperative Fish and Wildlife Research Unit \\ Department of Fish, Wildlife, and Conservation Biology \\ Department of Statistics \\ Colorado State University\\
    and \\
    Devin S. Johnson \\
    Alaska Fisheries Science Center \\ National Marine Fisheries Service \\ National Oceanic and Atmospheric Administration}
  \maketitle
} \fi

\if0\blind
{
  \bigskip
  \bigskip
  \bigskip
  \begin{center}
    {\LARGE\bf Basis Function Models for Animal Movement}
\end{center}
  \medskip
} \fi

\bigskip
\begin{abstract}
Advances in satellite-based data collection techniques have served as a catalyst for new statistical methodology to analyze these data. In wildlife ecological studies, satellite-based data and methodology have provided a wealth of information about animal space use and the investigation of individual-based animal-environment relationships.  With the technology for data collection improving dramatically over time, we are left with massive archives of historical animal telemetry data of varying quality. While many contemporary statistical approaches for inferring movement behavior are specified in discrete time, we develop a flexible continuous-time stochastic integral equation framework that is amenable to reduced-rank second-order covariance parameterizations. We demonstrate how the associated first-order basis functions can be constructed to mimic behavioral characteristics in realistic trajectory processes using telemetry data from mule deer and mountain lion individuals in western North America. Our approach is parallelizable and provides inference for heterogeneous trajectories using nonstationary spatial modeling techniques that are feasible for large telemetry data sets.
\end{abstract}

\noindent%
\vfill

\newpage
\spacingset{1.45} 

\section{Introduction}
Advancements in satellite data collection techniques have stimulated the development of dynamic statistical models for individual-based movement processes (\citealt{Kays:15}).  Individual-based statistical movement models have been used in a variety of recent applications including: vehicle (e.g., \citealt{Gloaguen:15}), cellular phone (e.g., \citealt{Calabrese:11}), and wildlife (e.g., \citealt{Hooten:10a}) tracking.   In particular, new inferential tools are crucial for improving the understanding of wildlife behavior and the response of individual animals to changing landscapes and environmental conditions.   Modern telemetry technology allows for remote data collection via ``on board'' devices (e.g., often using satellite-based observations of geographic position) and has provided massive repositories of information (e.g., \citealt{WikelskiKays:15}).   

For example, our application is focused on inferring the movement dynamics of two species of animals, mule deer (\emph{Odocoileus hemionus}) and mountain lion (\emph{Puma concolor}), using archival data sources.  Both species occur in western North America and our satellite telemetry data arose from two separate studies to investigate animal spatial ecology where tracking devices were fitted to individuals and recovered later.   

Despite numerous improvements to data collection methodology for remote tracking of individual animals, several remaining features of contemporary telemetry data must be addressed when making statistical inference.  For example, all forms of remotely collected telemetry data (i.e., measured geographic locations or positions) are susceptible to measurement error that can depend on the device, satellite system, terrain, land cover, weather, and behavior.  Recent advances have led to improved data modeling techniques that properly incorporate (and sometimes estimate) the uncertainty associated with telemetry measurement error (e.g., \citealt{Brost:15}; \citealt{Buderman:16}; \citealt{McClintock:15}).  

Irregular temporal measurement is another important feature to consider when modeling telemetry data.  Telemetry devices are often programmed (i.e., duty cycled) to record position data at a pre-specified set of times.  However, the frequency and regularity of these times are not consistent across studies.  Furthermore, despite the deterministic programming of satellite telemetry devices, missing data can occur stochastically due to instrumental difficulties as well as environmental and behavioral influences (e.g., terrain and weather).      

Dynamic statistical models for animal movement that formally incorporate measurement error typically assume a hierarchical structure (\citealt{Berliner:96}) where $\mathbf{s}(t_i)$ are the measured positions at time $t_i$ (for observation $i=1,\ldots,n$) and depend on the true positions $\boldsymbol\mu(t_i)$, that arise as a dynamical process.  Various process models have been proposed for the true underlying individual positions $\boldsymbol\mu(t_i)$, depending on the desired form of inference (\citealt{Hooten:17}).  Primarily, statistical models for position processes have fallen into three main categories: 1.) point process models (e.g., \citealt{Johnson:08a}; \citealt{Forester:09}; \citealt{Johnson:13}; \citealt{Brost:15}), 2.) discrete-time dynamic models (e.g., \citealt{Morales:04}; \citealt{Jonsen:05}; \citealt{McClintock:12}), and 3.) continuous-time dynamic models (e.g., \citealt{DunnGipson:77}; \citealt{Blackwell:97}; \citealt{Brillinger:01}; \citealt{Johnson:08b}; \citealt{Brillinger:10}).  While there have been similar advancements in each of these classes of movement models, we focus on the continuous-time formulations in what follows.  

Despite the popularity of point process models and discrete-time dynamic models, both present computational difficulties that prohibit widespread use for all but the simplest forms.  Point process models require numerical integration to calculate the likelihood (\citealt{Cressie:93}; \citealt{BermanTurner:92}; \citealt{WartonShepherd:10}; \citealt{Aarts:12}).  Discrete-time dynamic models often lack a joint model specification and require iterative calculation (\citealt{Morales:04}; \citealt{Jonsen:05}; \citealt{McClintock:12}).  Furthermore, discrete-time models provide inference relative to the scale of temporal discretization and must contain some mechanism to reconcile the times at which data are available with those of the latent discrete-time process (\citealt{McClintock:12}).
            
Discrete-time dynamic models for telemetry data are popular because they are heuristically straightforward to understand.  They have a long history of use and there is a large body of existing literature associated with modeling discrete time series (e.g., \citealt{Anderson-SprecherLedolter:91}).  Discrete-time models can also be extended to incorporate change-points and hidden Markov processes that allow for time-varying changes in the dynamics, hence better accommodating heterogeneous animal behavior through time (e.g., \citealt{Morales:04}).    

Given the attractive properties of discrete-time formulations for model building and the continuous-time nature of the true underlying trajectory, we present a general framework for constructing continuous-time models for animal movement based on limiting processes involving discrete-time models.  Our approach provides an intuitive connection to previously existing continuous-time stochastic process models for telemetry data (e.g., \citealt{DunnGipson:77}; \citealt{Blackwell:97}; \citealt{Brillinger:01}; \citealt{Johnson:08b}; \citealt{Brillinger:10}).  Borrowing techniques commonly used in spatial and spatio-temporal statistics (\citealt{CressieWikle:11}), we show how continuous-time models based on first-order (i.e., mean) dynamic structure can be implemented using equivalent second-order (i.e., covariance) specifications.  Second-order model specifications for animal movement applications have appeared only recently in the literature (e.g., \citealt{Fleming:14}; \citealt{Fleming:16}).  

We present a natural basis function approach to constructing appropriate covariance models for movement processes.  Our basis function specification is amenable to rank reduction and, thus, is computationally feasible to implement for large telemetry data sets.  We also demonstrate how the choice of basis function may correspond to an explicit representation of animal cognitive processes which may involve memory and perception.  Finally, to allow for heterogeneous dynamics in movement, we induce a non-stationary continuous-time temporal process by appropriately warping the temporal domain (i.e., temporal deformation, \citealt{SampsonGuttorp:92}).  Our warping approach allows for temporal clustering of movement behavior in continuous time similar to popular state-switching approaches in discrete-time models (e.g., \citealt{Morales:04}; \citealt{Hanks:11}; \citealt{McClintock:12}).  We also describe a parallelization strategy that can reduce required computational time substantially.    

\section{Methods}
\subsection{Continuous-Time Stochastic Trajectory Models}
We focus on process models for continuous-time trajectories (e.g., individual animal movement) and begin by describing dynamic model specifications for trajectories in multiple dimensions.  We return to measurement error models for observed trajectories in the next section.  

Many hierarchical statistical models involving continuous-time processes have relied on direct connections to Eulerian differential equations (e.g., \citealt{CangelosiHooten:09}).  In contrast, we begin with a discrete-time representation and describe the trajectory of a moving particle in terms of its position $\boldsymbol\mu(t_i)$, at time $t_i$, as   
\begin{align}
  \boldsymbol\mu(t_i) &= \boldsymbol\mu(t_0) + \lim_{\Delta t \rightarrow 0}\sum_{j=1}^i \boldsymbol\varepsilon(t_j) \label{eq:mut_brownian1} \\ 
  &= \boldsymbol\mu(t_0) + \mathbf{b}(t_i) \; ,
  \label{eq:mut_brownian2}
\end{align}
\noindent where, $\boldsymbol\mu(t_0)$ represents the beginning position at time $t_0$, and the limiting sum in (\ref{eq:mut_brownian1}) accumulates a sequence of discrete steps (i.e., displacement vectors, $\boldsymbol\varepsilon(t_j)$) up to time $t_i$.  If the displacement vectors, $\boldsymbol\varepsilon(t_j)\equiv \boldsymbol\mu(t_j)-\boldsymbol\mu(t_{j-1})$, are independent multivariate Gaussian, such that $\boldsymbol\varepsilon(t_j)\sim \text{N}(\mathbf{0},\sigma^2 \Delta t \mathbf{I})$ (i.e., white noise), we arrive at a continuous-time stochastic integral representation for $\boldsymbol\mu(t_i)$ where $\sigma$ controls the magnitude of displacement and $\mathbf{b}(t_i)$ is scaled multivariate Brownian motion (i.e., a multivariate Weiner process).  In the limiting trajectory model (\ref{eq:mut_brownian1}), the number of displacement steps in the sum increases as $\Delta t \rightarrow 0$, resulting in an infinite series that is often written as the integral 
\begin{equation}
  \mathbf{b}(t_i) = \int_{t_0}^{t_i} d\mathbf{b}(\tau)  \; , 
  \label{eq:bt_ito}
\end{equation}
\noindent using Ito notation (\citealt{Protter:04}). 

While the Brownian motion (BM) process model for an individual animal trajectory in (\ref{eq:bt_ito}) is a stochastic integral equation (SIE), it is also common to see it expressed as a stochastic differential equation (SDE) by differentiating both sides of (\ref{eq:mut_brownian1}) using Ito calculus (e.g., \citealt{DunnGipson:77}, \citealt{Blackwell:97}, \citealt{Brillinger:01}, and \citealt{Preisler:04}).   

\subsection{Smoothness in Trajectory Models}
\begin{figure}[htp]
  \centering
  \includegraphics[width=5in, angle=0]{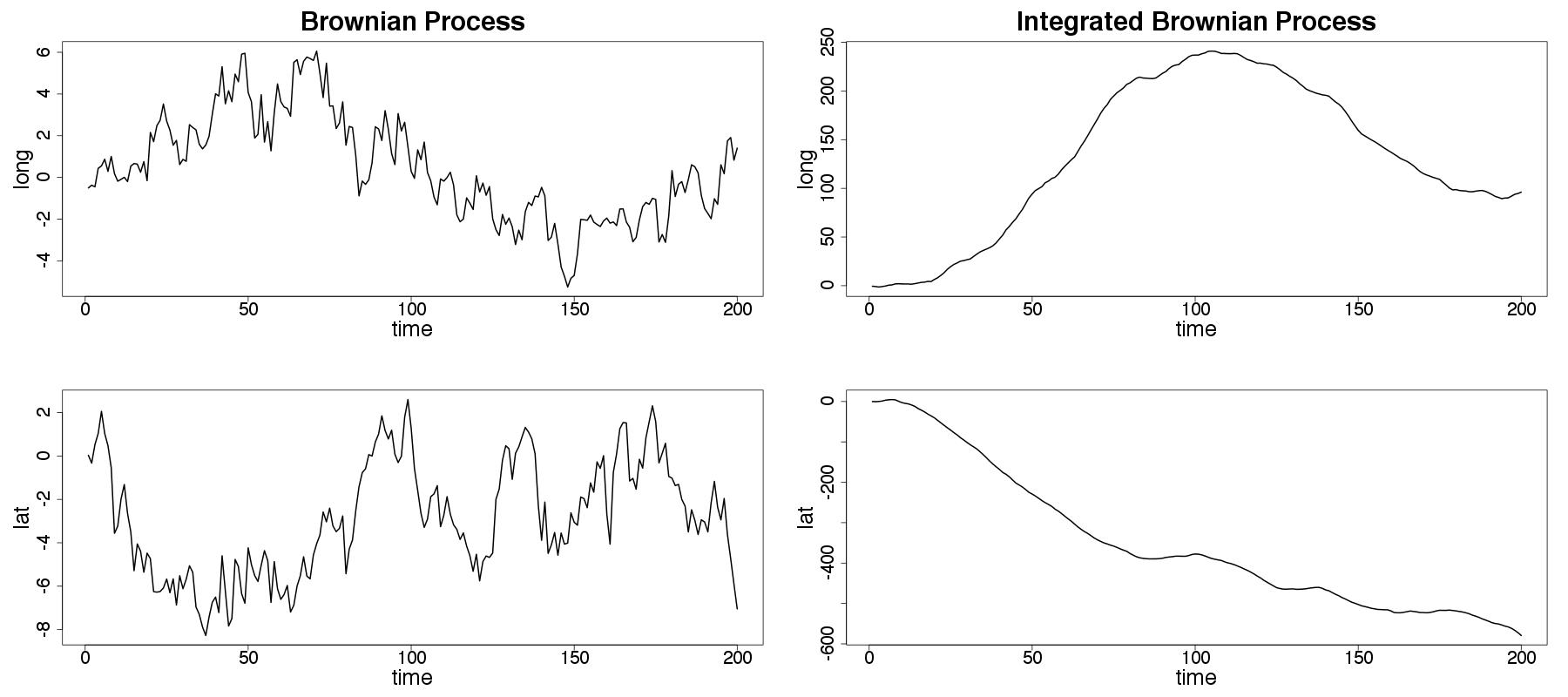}
  \caption{Left panels: Two-dimensional Brownian motion process, $\mathbf{b}(t)$.  Right panels:  Two-dimensional integrated Brownian motion, $\boldsymbol\eta(t)$.  Processes are displayed marginally in longitude and latitude.}
  \label{fig:bm_vs_dj}
\end{figure}

Stochastic trajectory models based on Brownian motion (e.g., (\ref{eq:mut_brownian2})) are inherently noisy (Figure~\ref{fig:bm_vs_dj}, left panels) as continuous-time processes.  \cite{Johnson:08b} presented a stochastic trajectory model for the velocity associated with animal movement that was integrated over time to yield a smoother position process (Figure~\ref{fig:bm_vs_dj}, right panels).  

We show that the velocity model of \cite{Johnson:08b} fits into a larger class of stochastic trajectory models by reparameterizing the simple Brownian SIE (\ref{eq:mut_brownian2}).  Recall that Brownian motion ($\mathbf{b}(t)$), at time $t$, can be expressed as an integral of white noise.  Thus, if we integrate Brownian motion itself, with respect to time, we have 
\begin{equation}
  \boldsymbol\eta(t) = \int_{t_0}^t \mathbf{b}(\tau) d\tau  \; ,  
  \label{eq:intBM}
\end{equation}
\noindent where $\boldsymbol\eta(t)$ is a version of the integrated stochastic process proposed by \cite{Johnson:08b}. This integrated Brownian motion (IBM; e.g., \citealt{Shepp:66}; \citealt{WeckerAnsley:83}; \citealt{RueHeld:05}) model (\ref{eq:intBM}) can be likened to that of \cite{Johnson:08b} by substituting $\boldsymbol\eta(t)$ into the position process (\ref{eq:mut_brownian2}) to yield $\boldsymbol\mu(t) = \boldsymbol\mu(t_0) + \boldsymbol\eta(t)$.  \cite{Jonsen:05} set an earlier precedent for modeling velocity, but strictly in a discrete-time framework.  

The approach proposed by \cite{Johnson:08b} suggests a more general framework for modeling movement that can be obtained by reparameterizing the velocity model using the $2\times 2$ matrix $\mathbf{H}(t,\tau)$ with diagonal elements equal to the function 
\begin{equation}
  h(t,\tau) = 
  \begin{cases}
    1  &\mbox{if } t_0 < \tau \leq t \\
    0  &\mbox{if } t < \tau \leq t_n  
  \end{cases} \; , 
  \label{eq:hJohnson}
\end{equation}
\noindent where $t_n$ is the last time at which data are observed, and off-diagonal elements equal zero.  Substituting $\mathbf{H}(t,\tau)$ into (\ref{eq:intBM}), the velocity-based Brownian motion model appears as the convolution 
\begin{equation}
  \boldsymbol\eta(t) = \int_{t_0}^{t_n} \mathbf{H}(t,\tau) \mathbf{b}(\tau) d\tau  \; .  
  \label{eq:intBMconv}
\end{equation}
\noindent The convolution in (\ref{eq:intBMconv}) is the key feature in a more general class of stochastic process models for animal movement trajectories.  For example, if $h(t,\tau)$ is a continuous function, such that $t_0 \leq t \leq t_n$, $t_0 \leq \tau \leq t_n$ and with finite positive integral $0 < \int_{t_0}^{t_n} h(t,\tau) d\tau < \infty$, then a new general class of continuous-time animal movement models arises.  We refer to this class of models as ``functional movement models'' (FMMs; \citealt{Buderman:16}). 

The ability to specify continuous-time movement models as a convolution (\ref{eq:intBMconv}) has two major advantages.  First, it clearly identifies the connections among animal movement models and similar models used in spatial statistics and time series.  Second, for the same reasons convolution specifications are popular in spatial statistics and time series, we show that FMMs share similar advantageous properties.   

In Appendix A (Supplementary Material), we show that the FMM in (\ref{eq:intBMconv}) can be rewritten as:  
\begin{equation}
  \boldsymbol\eta(t) = \int_{t_0}^{t_n} \tilde{\mathbf{H}}(t,\tau) d\mathbf{b}(\tau) \; , 
  \label{eq:intHtilda} 
\end{equation}
\noindent where $\tilde{\mathbf{H}}(t,\tau)$ has diagonal elements  
\begin{equation}
  \tilde{h}(t,\tau) = \int_{\tau}^{t_n} h(t,\tilde\tau) d\tilde\tau  \; .
  \label{eq:htilda}
\end{equation}
The FMM in (\ref{eq:intHtilda}) has the same form described in spatial statistics as a ``process convolution'' (or kernel convolution; e.g., \citealt{BarryVerHoef:96}; \citealt{Higdon:98}; \citealt{Lee:05}; \citealt{Calder:07}).  The process convolution has been instrumental in many fields, but especially in spatial statistics for allowing both complicated and efficient representations of covariance structure.  

\begin{figure}[htp]
  \centering
  \includegraphics[height=7in, angle=0]{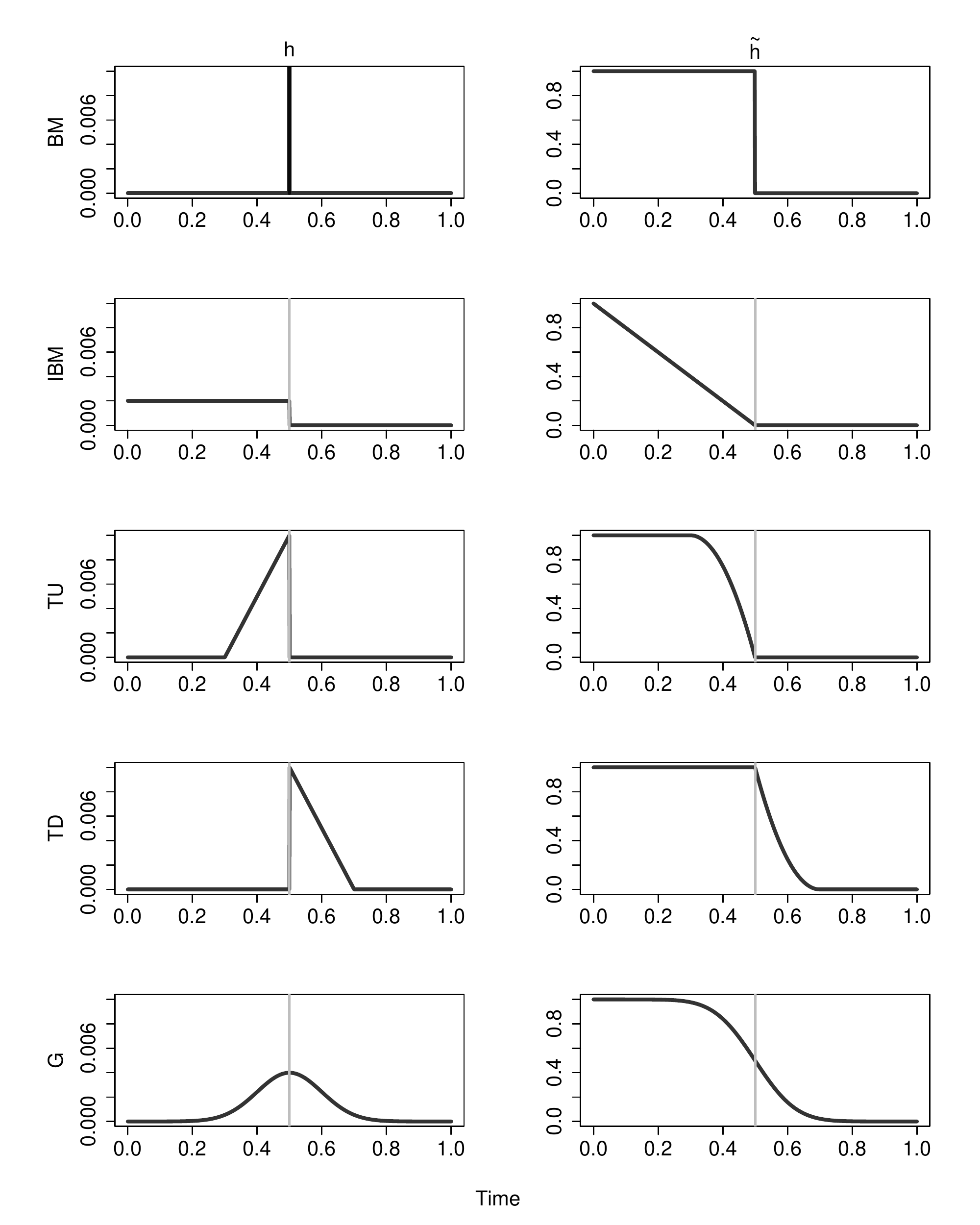}
  \caption{Example kernels $h(t,\tau)$ (left column) and resulting integrated kernels $\tilde{h}(t,\tau)$ (right column). The first row results in the regular Brownian motion (BM), row two shows integrated Brownian motion (IBM), rows three and four show tail-up (TU) and tail-down (TD) kernel functions, and row five shows the Gaussian (G) kernel functions.  Rows three through five (TU, TD, and G) are more common in time series and spatial statistics. The vertical gray line indicates time $t$ for the particular kernel shown; in this case $t=0.5$.}
  \label{fig:hforms}
\end{figure}

It is clear from (\ref{eq:intHtilda}) that we need not simulate Brownian motion, rather, we can operate on it implicitly by transforming the matrix function $\mathbf{H}(t,\tau)$ to $\tilde{\mathbf{H}}(t,\tau)$ via integration and convolving $\tilde{\mathbf{H}}(t,\tau)$ with white noise directly.  For example, consider the Gaussian kernel as the function $h(t,\tau)$.  The Gaussian kernel is one of the most commonly used functions in kernel convolution methods (e.g., \citealt{BarryVerHoef:96}; \citealt{Higdon:98}).  We convert $h(t,\tau)$ to the required function $\tilde{h}(t,\tau)$ using (\ref{eq:htilda}) and arrive at a numerical solution for the new kernel function by subtracting the normal cumulative distribution function (CDF) from one, a trivial calculation in most statistical software.  With respect to the time domain, the kernel $\tilde{h}(t,\tau)$ appears different than most kernels used in time series or spatial statistics (Figure~\ref{fig:hforms}, G, row 5).  Rather than being unimodal and symmetric, it has a sigmoidal shape equal to one at $t=t_0$ and nonlinearly decreasing to zero at $t=t_n$ resembling an I-spline (\citealt{Ramsay:88}).  In effect, this new kernel is accumulating the white noise up to near time $t$ and including a discounted amount of white noise ahead of time $t$.  

There are many options for kernel functions and each will result in different stochastic process models for animal movement.  In fact, we have already shown that the FMM class of movement models is general enough to include an integrated Brownian motion model similar to that developed by \cite{Johnson:08b}.  The FMM class also includes the original unsmoothed Brownian motion process if we let $h(t,\tau)$ be a point mass function at $\tau=t$ and zero elsewhere (Figure~\ref{fig:hforms}).  An approximation to the point mass kernel function can also be achieved by taking the limit as $\phi\rightarrow0$ of our Gaussian kernel and results in an integrated kernel function of   
\begin{equation}
  \tilde{h}(t,\tau) = 
  \begin{cases}
    1  &\mbox{if } t_0 < \tau \leq t \\
    0  &\mbox{if } t < \tau \leq t_n  
  \end{cases} \; . 
  \label{eq:htildaBM}
\end{equation}
\noindent The integrated kernel sums all past velocities to obtain the current position. The steep drop at $\tau=t$ is what induces roughness in the original Brownian motion process (Figure~\ref{fig:hforms}, BM, row 1).  Whereas, when we use a non-pointmass function for $h(t,\tau)$, we arrive at a smoother stochastic process model for movement.   

We highlight a few different kernel functions to examine their implications for animal movement behavior.  In doing so, it is simplest to interpret the $h(t,\tau)$ and $\tilde{h}(t,\tau)$ functions directly.  For example, using the direct integration of velocity as proposed by \cite{Johnson:08b} results in the integrated Brownian motion (IBM) kernel ($\tilde{h}(t,\tau)$) on the second row in Figure~\ref{fig:hforms}.  In this case, the individual's position accumulates its past steps, which are noisy themselves in direction and length, but have some general momentum.  The ``tail up'' (TU) kernel shown in the third row of Figure~\ref{fig:hforms} models the current position based on past steps that decay linearly with time (we borrow the tail up and tail down terminology from \citealt{VerHoefPeterson:10}).  In this case, the individual's position is more strongly a function of recent steps than steps in the distant past.  The opposite is true with the ``tail down'' (TD) kernel shown in row 4 of Figure~\ref{fig:hforms}, where only future steps influence position.  Heuristically, we interpret the resulting movement as perception driven.  That is, the individual may have an awareness of a distant destination that affects their movement.  Finally, the Gaussian kernel discussed earlier and shown in the bottom row of Figure~\ref{fig:hforms} indicates a symmetric mixture of previous and future velocities, suggesting an equal perception of former and future events by the individual.  Appendix B (Supplementary Material) contains the functional forms and a visualization of the basis functions in $\tilde{\mathbf{H}}$.       

\subsection{Functional movement models and covariance}
In the statistical setting, we seek inference for model parameters as well as predictions (interpolation) of the position process in time.  Therefore, we need to connect telemetry data at a finite set of observation times \{$t_1,\ldots,t_n$\} to the dependence structure imposed by the FMM.  A common approach for expressing dependence structure in correlated processes is using covariance.  Translating dynamically induced temporal structure into covariance yields several computational advantages for FMMs.  Ultimately, we construct a hierarchical statistical model for telemetry data $\mathbf{s}(t_i)$, for $i=1,\ldots,n$, that relies on a latent dynamical process governed by an FMM.  In what follows, we describe the covariance properties of FMMs and then specify a hierarchical model that exploits those properties in the next Sections.     

Using the basic FMM in (\ref{eq:intBMconv}), we can choose from a large set of possible smoothing kernels ($h(t,\tau)$) for Brownian motion and arrive at the appropriate form for the integrated kernel ($\tilde{h}(t,\tau)$) that is convolved with white noise.  We can then use $\tilde{h}(t,\tau)$ directly to construct the proper covariance function for the joint process.  In fact, for a one-dimensional movement process $\eta(t)$, the covariance function can be calculated (e.g., \citealt{PaciorekSchervish:06}) as the convolution of the kernels 
\begin{equation}
  \text{cov}(\eta(t_1),\eta(t_2))=\int_{t_0}^{t_n} \sigma^2 \tilde{h}(t_1,\tau)\tilde{h}(t_2,\tau)d\tau \; ,
  \label{eq:BMconvcov}
\end{equation}
\noindent for any two times, say $t_1$ and $t_2$.  

One benefit of the covariance function (\ref{eq:BMconvcov}) is that, for a finite subset of $n$ times $\{t_1, \ldots, t_n\}$ and process $\boldsymbol\eta \equiv (\eta_1,\ldots,\eta_n)'$, the joint probability model can be expressed as    
\begin{equation}
  \boldsymbol\eta\sim\text{N}(\mathbf{0},\sigma^2\Delta t \tilde{\mathbf{H}}\tilde{\mathbf{H}}') \; ,
  \label{eq:BMjoint}
\end{equation}
\noindent where $\mathbf{0}$ is an $n\times 1$ vector of zeros and $\tilde{\mathbf{H}}$ is a matrix of basis functions with the $i$th row equal to $\tilde{h}(t_i,\tau)$ for all $\tau$.  This procedure for constructing the covariance matrix and defining a correlated process is similar to that recommended in spatial statistics (e.g., \citealt{PaciorekSchervish:06}; \citealt{VerHoefPeterson:10}).  For simplicity, the resulting model for the joint one-dimensional position process ($\boldsymbol\mu$, an $n\times 1$ vector) at the observation times for smooth Brownian motion is   
\begin{equation}
  \boldsymbol\mu\sim\text{N}(\mu(0)\mathbf{1},\sigma^2\Delta t \tilde{\mathbf{H}}\tilde{\mathbf{H}}') \; . 
  \label{eq:BMjointmu}
\end{equation}

Despite the fact that it is often more intuitive to model the process from the first-moment (i.e., mean dynamical structure) rather than the second-moment (\citealt{WikleHooten:10}), the joint form of (\ref{eq:BMjointmu}) with dependence imposed through the $n\times n$ correlation matrix $\tilde{\mathbf{H}}\tilde{\mathbf{H}}'$ can be useful computationally (\citealt{Sampson:10}).  When the integral in (\ref{eq:BMconvcov}) cannot be used to analytically compute the necessary covariance matrix, we can still use the outer product of the matrices explicitly (i.e., $\tilde{\mathbf{H}}\tilde{\mathbf{H}}'$).  However, the true covariance requires the number of columns of $\tilde{\mathbf{H}}$ to approach infinity, which, under approximation, can lead to computational difficulties.  \cite{Higdon:02} suggested a finite process convolution as an approximation.  In the finite approximation, the number of columns of $\tilde{\mathbf{H}}$ could be reduced to say, $m$ columns.  This rank reduction implies that there are $m$ knots in the temporal domain that anchor the basis functions (i.e., kernels) and, thus, only $m$ white noise terms are required so that $\boldsymbol\eta \approx \tilde{\mathbf{H}}\boldsymbol\varepsilon$, where $\tilde{\mathbf{H}}$ is an $n\times m$ matrix and $\boldsymbol\varepsilon\equiv (\varepsilon(t_1),\ldots,\varepsilon(t_j),\ldots,\varepsilon(t_m))'$ is an $m\times 1$ vector.  The use of a finite approximation to the convolution is also sometimes referred to as a reduced-rank method (\citealt{Wikle:10}).  Rank reduction can improve computational efficiency and has become very popular in spatial and spatio-temporal statistics for large data sets (\citealt{NychkaSaltzman:98}; \citealt{Nychka:02}; \citealt{Banerjee:08};  \citealt{CressieWikle:11}).          

To illustrate how kernel functions relate to covariance thus far, we simplified the movement process so that it is one-dimensional in space.  The same approach generalizes to higher dimensions.  In the more typical two-dimensional case, we stack the vectors in each dimension to form a single $2n\times 1$ vector $\boldsymbol\eta$.  Then the joint model can be written as 
\begin{equation}
  \boldsymbol\eta\sim\text{N}(\mathbf{0},\sigma^2\Delta t (\mathbf{I}\otimes \tilde{\mathbf{H}}\tilde{\mathbf{H}}')) \; ,
  \label{eq:BMjointMV}
\end{equation}
\noindent where $\mathbf{0}$ is a $2n\times 1$ vector of zeros and $\mathbf{I}$ is a $2\times 2$ identity matrix (assuming that $\tilde{\mathbf{H}}$ is the appropriate set of basis functions for both directions; i.e., latitude and longitude).  

\subsection{Heterogeneous Dynamics}
Discrete-time trajectory models for animal movement (e.g., \citealt{Morales:04}) are commonly specified with time-varying dynamics.  In spatial statistics, heterogeneity in dependence structure is often treated as nonstationarity. A variety of approaches have been suggested for modeling nonstationary continuous processes (e.g., \citealt{SampsonGuttorp:92}; \citealt{Higdon:02}).  We describe a temporal warping approach to modeling nonstationary movement.  

Our warping approach (also known as ``deformation'') was described in a continuous spatial modeling context by \cite{SampsonGuttorp:92} and later extended (e.g., \citealt{Damian:01}; \citealt{SchmidtOHagan:03}).  In the warping approach, we map the times $\mathbf{t}\equiv (t_1,\ldots,t_n)'$ to a new set of times $\mathbf{w}\equiv(w_1,\ldots,w_n)'$ using a smooth function so that no ``folding'' occurs (i.e., transformed times retain order).  A model-based method to perform the warping allows $\mathbf{w}$ to arise as a correlated random field anchored by $\mathbf{t}$.  For example, a simple additive warping can be induced through the Gaussian process model $\mathbf{w} \sim \text{N}(\mathbf{t},\boldsymbol\Sigma_w)$, where a smooth covariance function, such as the Gaussian $\Sigma_{w,ij} \equiv \sigma_w^2 \exp(-(t_i - t_j)^2/\phi_w^2)$, is chosen.  Conventionally, the warping parameters $\sigma_w^2$ and $\phi_w$ are constrained to yield non-folding temporal warp fields. 

When fitting the FMM (\ref{eq:BMjointmu}), we construct the basis functions so that they depend on the warped times.  For example, using a Gaussian kernel, we now have 
\begin{equation}
  h(t,\tau) \propto \exp \left(-\frac{(w(t)-\tau)^2}{\phi^2} \right) \; . 
\end{equation}
\noindent Notice that the kernel $h(t,\tau)$ only depends on a single range parameter $\phi$ that does not need to vary in time because the heterogeneity enters through the norm in the warped temporal domain $(w(t)-\tau)^2$.  Thus, the warp field $\mathbf{w}$ is a latent process to be estimated in the model and can provide inference associated with changes in movement dynamics.  For example, the derivative $dw(t)/dt$ of the warp field provides insight about changes in the individual's velocity over time.  When $dw(t)/dt$ is large, the warp field expands the temporal domain to allow for smoother paths, either from faster, directed movement or lack of movement.  Expanding the temporal domain can account for behavior indicative of long-distance migration or sedentariness.  By contrast, when the $dw(t)/dt$ is small, the temporal domain compresses, resulting in more tortuous paths and accounting for periods of time with sharper turning angles.  Temporal compression leads to behavior more typical of an area restricted search (e.g., \citealt{KnellCodling:12}), or an individual performing routine activities within its home range (e.g., foraging). 

\subsection{Full Model Specification}
Assuming Gaussian measurement error associated with the telemetry observations and a full-rank position process $\boldsymbol\mu$, we specify the full model in a hierarchical Bayesian framework:   
\begin{align}
  \mathbf{s} &\sim \text{N}(\mathbf{K}\boldsymbol\mu,\mathbf{I}\otimes \boldsymbol\Sigma_s) \;, \label{eq:datamod} \\
  \boldsymbol\mu &= \boldsymbol\mu(0)\otimes \mathbf{1} + (\mathbf{I}\otimes \tilde{\mathbf{H}})\boldsymbol\varepsilon \;, \\ 
  \boldsymbol\varepsilon &\sim \text{N}(\mathbf{0},\sigma^2\Delta t \mathbf{I}) \; , \label{eq:epsmod} 
\end{align}
\noindent where $\mathbf{K}$ is a $2n\times 2m$ matrix that maps the data to the underlying process at the appropriate times and the error covariance matrix could be left general or simplified as $\boldsymbol\Sigma_s\equiv \sigma^2_s\mathbf{I}$ depending on the data source.  As before, the initial position is denoted as $\boldsymbol\mu(0)$ and the process variance is $\sigma^2$.  In the model formulation that allows for temporal heterogeneity, the matrix of basis functions depends on the range parameter $\phi$, as well as the warped time process $\mathbf{w}$.  Thus, we augment the model specification by letting $\mathbf{w} \sim \text{N}(\mathbf{t},\boldsymbol\Sigma_w)$.        

In principle, the choice of priors is study specific, but any proper prior distributions with the correct support may be used for the parameters in the FMM.  In Gaussian state-space models, conjugate priors for $\sigma^2_s$ and $\sigma^2$ are inverse gamma, where strong prior information helps to separate the measurement error variance $\sigma^2_s$.  Thus, similar to geostatistics, inference can be sensitive to the chosen prior (e.g., \citealt{Finley:15}) when available information is used to inform priors.  Alternative priors for the latent process variance are also available.  For example, we used a proper uniform prior for $\sigma$, as recommended by \cite{Gelman:06} for latent Gaussian processes.  A variety of options are available for the temporal range parameter $\phi$, however, a discrete uniform prior ($\phi \sim \text{DiscUnif}(\boldsymbol\Phi)$) is computationally advantageous (\citealt{DiggleRibeiro:02}).  We elaborate on the discrete uniform prior for the temporal range parameter ($\phi$) in the next Section.  

\subsection{Model Implementation}
The full model, described in the previous section, can be implemented using an MCMC algorithm to sample from all full-conditional distributions sequentially.  However, such an algorithm would be prohibitively slow for all but the smallest telemetry data sets.  Thus, we provide a model reparameterization that results in several computational improvements.  Our approach involves three critical features: 1.) Discrete uniform prior for the range parameter, 2.) integrated likelihood, and 3.) a mixture model to accommodate temporal heterogeneity.  We describe each of these features in what follows.

When specifying the prior $\phi \sim \text{DiscUnif}(\boldsymbol\Phi)$, for each value of $\phi$ in the support $\boldsymbol\Phi$, the entire matrix of basis functions $\tilde{\mathbf{H}}$ can be precomputed and accessed as needed in a Markov Chain Monte Carlo (MCMC) algorithm.  The discrete uniform prior permits an MCMC algorithm that requires only minimal matrix calculations (\citealt{DiggleRibeiro:02}).   

We used Rao-Blackwellization to derive an integrated likelihood that results in a more stable MCMC algorithm and allows us to avoid sampling the latent process directly.  The integrated likelihood provides better mixing for covariance parameters and the latent process can be recovered \emph{post hoc} using saved MCMC samples and composition sampling in a secondary algorithm.  This approach is a common computational strategy used in spatial statistics (e.g., \citealt{Finley:15}). In our case, multivariate normal properties allow us to integrate $\boldsymbol\varepsilon$ out of our hierarchical model, resulting in the integrated model formulation  
\begin{equation}
  \mathbf{s} \sim \text{N}(\mathbf{K}(\boldsymbol\mu(0)\otimes\mathbf{1}),\mathbf{I}\otimes \boldsymbol\Sigma_s + \sigma^2 \mathbf{K}(\mathbf{I}\otimes \tilde{\mathbf{H}})(\mathbf{I}\otimes \tilde{\mathbf{H}})'\mathbf{K}') \;.
  \label{eq:integratedmodel}
\end{equation}
\noindent Note that we omit the time step $\Delta t$ from the white noise specification because the variance term $\sigma^2$ can account for it in the discrete-time implementation.  The resulting combined covariance matrix in (\ref{eq:integratedmodel}) includes both the measurement error variance and the process covariance and has the form $\mathbf{A}+\mathbf{BCD}$, which can be inverted efficiently using the Sherman-Morrison-Woodbury identity if necessary (Appendix C, Supplementary Material).  In addition to the precalculation of terms involving the matrix of basis functions $\tilde{\mathbf{H}}$, fast matrix calculations are essential for fitting the FMM efficiently to real data sets.     

The two steps described above facilitate the fitting of a temporally homogeneous model that is conditioned on a known warping ($\mathbf{w}$).  However, because the optimal warping is unknown and occurs in a nonlinear covariance function for the data, we would need to sample it using Metropolis-Hastings within the broader MCMC algorithm for fitting the model.  The resulting computational requirements are prohibitively large.  Thus, we propose a mixture model framework that can be implemented using reversible-jump MCMC (RJMCMC; \citealt{Green:95}).  In practice, the mixture model can be implemented using a finite set of candidate temporal warp fields, say $\mathbf{w}_j$, for $j=1,\ldots,J$ and $L$ basis function types (Figure~\ref{fig:hforms}) resulting in $N=L\cdot J$ total models.  We formulate the mixture using models ${\cal M}_{lj}$ (for $l=1,\ldots,L$ and $j=1,\ldots,J$) and latent indicator variables $z_{lj}$ as
\begin{equation}
  \mathbf{s} \sim 
  \begin{cases}
    {\cal M}_{11} &\mbox{, }  z_{11} = 1 \\
    {\cal M}_{21} &\mbox{, }  z_{12} = 1 \\ 
    \vdots  \\
    {\cal M}_{LJ} &\mbox{, }  z_{LJ} = 1 \\ 
  \end{cases}
  \label{eq:mixturemodel}
\end{equation}
\noindent where ${\cal M}_{lj}\equiv \text{N}(\mathbf{K}(\boldsymbol\mu(0)\otimes\mathbf{1}),\mathbf{I}\otimes \boldsymbol\Sigma_s + \sigma^2 \mathbf{K}\mathbf{R}_{lj}(\phi,\mathbf{w}_j)\mathbf{K}')$ for $l=1,\ldots,L$ and $j=1,\ldots,J$, and each correlation matrix is defined as $\mathbf{R}_{lj}\equiv(\mathbf{I}\otimes \tilde{\mathbf{H}}_{lj})(\mathbf{I}\otimes \tilde{\mathbf{H}}_{lj})'$ (with implicit dependence on the appropriate basis function type, range parameter, and temporal warping).  The indicator variables $\mathbf{z}\equiv(z_{11},z_{12},\ldots,z_{LJ})'$ are multinomial ($\mathbf{z} \sim \text{MN}(1,\mathbf{p})$) with $\mathbf{p}$ serving as prior model probabilities.  Using the mixture model specification in (\ref{eq:mixturemodel}), and conditioning on a large set of potential warp fields, the resulting MCMC algorithm involves Gibbs updates for the indicator variables, $z_{lj}$.  The marginal posterior mean of the indicator variable $z_{lj}$, $E(z_{lj} | \mathbf{s})$, corresponds to the posterior model probability for model ${\cal M}_{lj}$, $P({\cal M}_{lj} | \mathbf{s})$.    

An alternative computational strategy involves the two-stage RJMCMC approach of \cite{BarkerLink:13} to find the posterior model probability $P({\cal M}_{lj} | \mathbf{s})$ and model averaged posterior inference (Appendix D, Supplementary Material).  The advantage of the RJMCMC approach of \cite{BarkerLink:13} is that each model can be fit independently (and in parallel), and a secondary MCMC algorithm can be used to sample from the averaged posterior distributions.   Parallel model fits provide appreciable gains in computational efficiency.  Finally, after fitting each of the $N$ models, and applying Bayesian model averaging via RJMCMC, we use a tertiary algorithm to obtain realizations of the latent position process $\boldsymbol\mu(t)$ for any time of interest $t$.  To fit the models we propose, the tertiary algorithm can also be parallelized, but is usually fast enough to implement sequentially.   

The two-stage RJMCMC facilitates the ability to compare models with different basis function specifications.  Thus, we include both, a set of candidate temporal warp fields and all of the covariance models constructed from the basis functions shown in Figure~\ref{fig:hforms}.  We accumulate the posterior model probability resulting from the RJMCMC algorithm for each form of basis function, resulting in $\sum_{j=1}^J P({\cal M}_{lj} | \mathbf{s})$ for each basis function type $l$, which may provide insight about potential mechanisms (i.e., perception and memory) that influence animal behavior and movement.  

Finally, to obtain a set of candidate warp fields, we simulate warp fields as Gaussian processes with Gaussian covariance structure to provide appropriate smoothness in the resulting temporal deformation.  High variance and low temporal correlation in the warp distribution results in a higher proportion of folding warp fields.  Thus, we use a rejection sampling algorithm to retain only those warp fields that obey the ordering of the original time domain (i.e., do not fold).  We simulate warp fields based on a range of parameter values to yield a candidate set that do not fold, yet contain a variety of multiscale patterns.  The appropriate temporal scale for the warp distribution is informed by the data through the model averaging procedure.               

In simulation, our method for fitting the FMM to data was approximately an order of magnitude faster than conventional Bayesian computational methods.  We demonstrate that our Bayesian FMM recovers model parameters for simulated data sets with varying amounts of data missingness and temporal heterogeneity in Appendix E (Supplementary Material).  In what follows, we apply the temporally heterogeneous FMM to satellite telemetry data corresponding to the movement of large mammals in western North America and describe the implications for our proposed methods.  

\section{Data Analysis: Animal Movement}
We apply our temporally heterogeneous FMM to satellite telemetry data sets consisting of recorded mule deer (\emph{Odocoileus hemionus}) and mountain lion (\emph{Puma concolor}) positions in Southeastern Utah and Northern Colorado, USA (originally analyzed for other purposes by \cite{Hooten:10a} and \cite{Hanks:15}).  Mule deer and mountain lion are important species in the western North America that often exhibit seasonally varying movement behavior (\citealt{Ager:03}; \citealt{Pierce:99}).  Many mule deer individuals split their time between winter and summer ranges and migrate from one to the other in the spring and autumn.  Mountain lion movement behavior also varies temporally, with individuals roaming regionally and often returning to locations where prey has been captured.  In both mule deer and mountain lion individuals, there is a clear heterogeneity in movement dynamics.  Our FMM approach accommodates such movement heterogeneity by treating it as temporal nonstationarity.  

For the mule deer example, we focus on a 5 day period during autumn of 2005 that spans the migratory behavior of an individual near Castle Valley, Utah, USA.  These data were collected as part of a larger study of the spread of chronic wasting disease (\citealt{McFarlane:07}), a contagious prion disease that occurs in North American ungulates (e.g., \citealt{Farnsworth:06}; \citealt{Miller:06} \citealt{Garlick:11}; \citealt{Garlick:14}; \citealt{Evans:16}).  Migratory routes are a critical element of the natural history of mule deer in western North America (\citealt{Nicholson:97}) and an improved understanding of them is essential for habitat management.     

For the mountain lion, we focus on a 33 day period during summer of 2011 in the Front Range of Colorado, USA.  These data were collected as part of a larger study to assess the movement and survival of mountain lions in the Front Range of Colorado.  The mountain lion is an important native carnivore that was once ubiquitous throughout much of North America (\citealt{YoungGoldman:46}; \citealt{Laundre:13}).  Mountain lions play an essential role in ecosystem function by attenuating populations of prey species, which often include mule deer (\citealt{Knopff:10}).  Inference concerning mountain lion movement is also important in the Front Range because of rapid development of exurban areas in existing habitat (\citealt{Blecha:15}).      

\begin{figure}[htp]
  \centering
  \vspace{-.75in}
  \includegraphics[height=6in, angle=0]{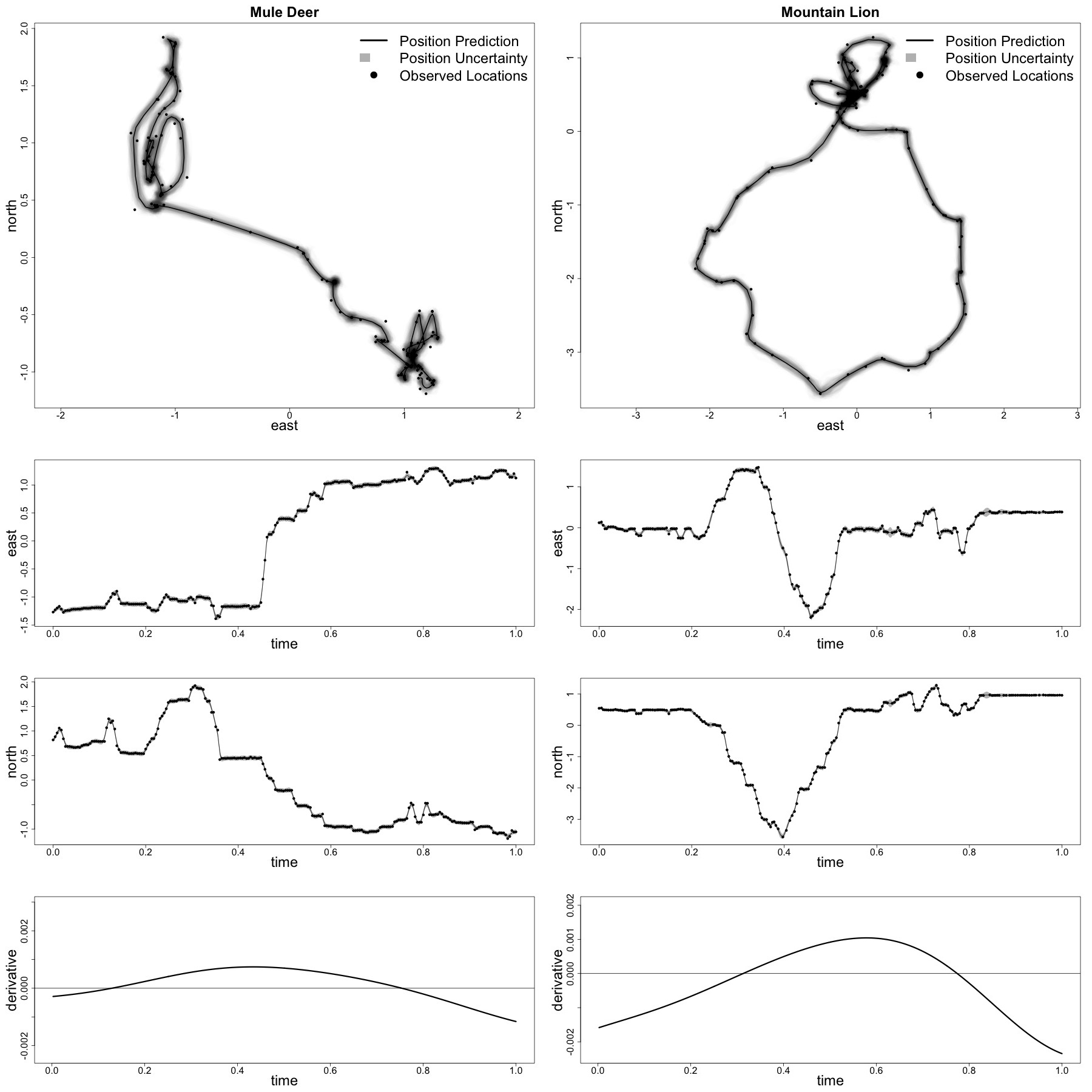}
  \caption{Posterior predicted mule deer (left column) and mountain lion (right column) paths $\boldsymbol\mu(t)$ using BMA.  Top panels:  Paths in two-dimensional geographic space.  Middle panels: Marginal paths for easting and northing.  Position uncertainty is represented as posterior realizations in the top panel and 95\% marginal credible intervals in the middle two panels. Bottom panel:  Warp derivative obtained via BMA.}
  \label{fig:MD_ML_path}
\end{figure}

The resulting telemetry data (Figure~\ref{fig:MD_ML_path}) for both species were obtained using global positioning system (GPS) telemetry devices affixed to the individuals and are comprised of 226 measured geographic positions spaced approximately 30 minutes apart for the mule deer individual and 221 observations spaced approximately 3 hours apart for the mountain lion individual.  For analysis and plotting purposes, we standardized the position data (i.e., subtracted the sample mean and divided by the pooled sample standard deviation) and converted the time scale to the [0,1] interval.  In principle, data transformation is not necessary for model implementation, but it allows us to use the same software to simulate warp fields, fit the models, as well as use similar prior specification among species, leading to fewer numerical overflow issues.    

To implement an FMM for the mule deer and mountain lion data using the specification presented in the previous section (\ref{eq:integratedmodel}), we used all of the kernel forms, $h_l(t,\tau)$ for $l=1,\ldots,5$, described in Figure~\ref{fig:hforms} (and Appendix B).  Following \cite{DiggleRibeiro:02}, we reparameterized the covariance matrix for a given basis function type $l$ in the FMM (\ref{eq:integratedmodel}) such that  
\begin{equation}
  \mathbf{I}\otimes \boldsymbol\Sigma_s + \sigma^2 (\mathbf{I}\otimes \tilde{\mathbf{H}}_l)(\mathbf{I}\otimes \tilde{\mathbf{H}}_l)' = \sigma^2_s (\mathbf{I} + \sigma^2_{\mu/s}(\mathbf{I}\otimes \tilde{\mathbf{H}}_l)(\mathbf{I}\otimes \tilde{\mathbf{H}}_l)'),  
\end{equation}
\noindent where, $\sigma^2_{\mu/s}=\sigma^2/\sigma^2_s$ and $\tilde{\mathbf{H}}_l(\phi)$ is a function of the temporal range parameter $\phi$ for the Gaussian, tail-up, and tail-down kernel forms.  For the standardized data, we used a discrete uniform prior for $\phi$ on 100 evenly spaced values from $0.001$ to $0.1$, a uniform prior for $\sigma_{\mu/s}$ on support $(0,20)$, and an informative inverse gamma prior for $\sigma^2_s \sim \text{IG}(12,0.01)$, representing strong prior information that GPS telemetry error is often less than 150 m.   

In implementing the FMM, we used 400 equally spaced temporal knots for the kernels and we sampled approximately 4000 candidate warp fields using a Latin hypercube design from a Gaussian process with Gaussian covariance constrained to avoid temporal folding.  The warp distribution was based on 100 parameter combinations for $\sigma_w$ and $\phi_w$ (10 equally spaced values for each parameter, ranging from 0.001 -- 1).  The resulting multiscale candidate warp fields represent a wide range of temporal deformation patterns that are combined optimally using Bayesian model averaging.  Trace plots for model parameters indicated excellent MCMC mixing and convergence, based on 10000 MCMC samples for each model fit.  Each model fit required only 3 minutes on a 16-core workstation with 3 Ghz processors and 64 GB of memory.  The total time required to fit the model for each of the warp fields in parallel was approximately 3 hours, which is substantially less than the 50 hours it would require to compute in sequence.  The secondary and tertiary algorithms to perform the BMA and posterior predictive sampling required only 4 additional minutes.  We used empirical variograms based on the posterior predictive residuals to check the models (Appendix F, Supplementary Material).  For both the mule deer and mountain lion data, the variograms showed no evidence of remaining temporal autocorrelation unaccounted for by the model.             

The resulting inference is summarized in Figure~\ref{fig:MD_ML_path}.  The top row of panels in Figure~\ref{fig:MD_ML_path} presents the data and estimated latent process for the mule deer and mountain lion individuals in two dimensional geographic space. The gray lines are posterior predictive realizations of the latent process $\boldsymbol\mu(t)$.  The middle two rows of panels in Figure~\ref{fig:MD_ML_path} display the marginal latent process in longitude and latitude.  The bottom row of panels in Figure~\ref{fig:MD_ML_path} shows the inferred derivative of the temporal warp at the posterior mean of the position process.      

The mule deer individual migrated during the time interval 0.3--0.6 (Figure~\ref{fig:MD_ML_path}).  As discussed in the Heterogeneous Dynamics Section, the warp derivative can provide insight on the movement dynamics.  The migratory movement behavior caused the BMA warp derivative to be positive, indicating that a temporal expansion is needed to accommodate the trajectory during that time period.  In contrast, the warp derivative was negative during the time interval 0.6--0.75 suggesting a temporal contraction was necessary to capture the different movement behavior for the individual directly after migration.  

The accumulated posterior model probabilities for the kernel forms suggested that Brownian motion and integrated Brownian motion were not helpful in describing mule deer movement based on these data (i.e., posterior model probability equal to zero).  However, the Gaussian, tail-up, and tail-down kernels accounted for 0.059\%, 0.920\%, and 0.021\%, respectively.  The distribution of posterior model probabilities for the kernel forms suggest that, while Gaussian, tail-up, and tail-down kernels have some non-zero contribution, the tail-up kernel dominates in characterizing mule deer movement.  Recall that the tail-up kernel acts as a local filter for Brownian motion, smoothing the individual trajectory using past Brownian steps.  A potential interpretation of the posterior model probabilities suggest that memory may play the largest role in mule deer movement during migration. 

The warp derivative for the mountain lion individual (Figure~\ref{fig:MD_ML_path}) suggests a temporal contraction at the beginning and end of the temporal domain (i.e., $t<0.2$ and $t>0.8$) with a temporal expansion between (i.e., $0.4 < t < 0.8$).   The FMM inference confirms that the mountain lion individual occupies a location with prey before traversing a large loop, only to return to the same location a few weeks later.  Mountain lion movement is thought to be strongly influenced by a knowledge of kill sites (\citealt{Hanks:15}) and our analysis provides additional quantitative evidence to support that.  

Similar to the mule deer results, the accumulated posterior model probabilities for the kernel forms suggested that Brownian motion and integrated Brownian motion were not helpful in describing mountain lion movement based on these data (i.e., posterior model probability equal to zero).  However, the Gaussian, tail-up, and tail-down kernels accounted for 0.27\%, 0.7295\%, and 0.0005\%, respectively.  The distribution of posterior model probabilities for the kernel forms suggest that, while both Gaussian and tail-up kernels are useful, the tail-up kernel is twice as important for characterizing mountain lion movement.  The combination of smoothing based on previous and future Brownian steps suggest that both memory and perception may influence mountain lion movement based on the data we analyzed.      

\section{Discussion}
The rapidly expanding literature on statistical models for animal movement has provided a wealth of useful tools for analyzing telemetry data.  In particular, individual-based models for trajectories have become a popular means for obtaining inference concerning the patterns and processes involved in animal movement.  Despite the development of approaches for modeling animal trajectories in continuous-time, discrete-time movement models have dominated the recent literature.  Heuristically, discrete-time animal movement models (e.g., \citealt{Morales:04}; \citealt{Jonsen:05}; \citealt{McClintock:12}) are straightforward and easy to understand.  They also provide a direct means to incorporate time-varying heterogeneity.  In particular, temporal clustering of animal movement trajectories has proven useful for inferring animal behavior and for better understanding the spatial patterning of behavior.  However, movement processes are necessarily continuous in time, but traditional continuous-time trajectory models have been oversimplified or computationally challenging to implement for large datasets (\citealt{McClintock:14}).    

We presented a continuous-time stochastic trajectory model that retains the natural intuition of discrete-time models, is feasible to implement for large data sets, accommodates time-varying heterogeneity in movement, and is general enough to allow for cognitive realism (e.g., memory and perception).  We showed that existing stochastic process models for movement can be generalized to facilitate basis function representations of the dynamical process.  Basis function representations are commonly used in spatial and spatio-temporal statistics, semi-parametric modeling, and functional data analysis.  We show how the same machinery used in spatial statistics to model complicated dependence structures in data can also be used in animal movement modeling.  For example, temporal heterogeneity can be expressed as non-stationarity in a second-order Gaussian process model for individual trajectories using existing methods.  

In previous implementations of both continuous- and discrete-time animal movement models, computation has been a bottleneck for inference based on statistical modeling.  Computationally feasible model structures were often over-simplified and lacked appropriate realism.  Historically, more complicated models required custom software written in compiled languages that do not facilitate user modifications.  We combined multiple computational approaches to develop a strategy for fitting hierarchical stochastic process models to telemetry data.  Our method exploits the substantially improved storage and multicore processing capabilities of modern computers so that the models we presented herein can be fit on a standard laptop computer. 

We showed that our modeling framework appropriately accommodates temporally heterogeneous structure in realistic animal trajectories as well as providing a natural way to handle irregular fix rates common in telemetry data.  Most discrete-time trajectory models for telemetry data rely on either imputation or a linear interpolation to accommodate irregular acquisition times.  For mule deer and mountain lion individuals, we showed that our model is able to accommodate mechanistic forms of temporal heterogeneity due to natural seasonality and environmental cues, providing inference concerning where and when changes in movement behavior occur.  

The basis function approach we present is both familiar and further generalizable.  For example, the Gaussian data and process model components in the FMM are important for computational tractability, however they can easily be generalized.  In particular, other forms of satellite telemetry data (e.g., Argos) have distinctly non-Gaussian telemetry error patterns (\citealt{Brost:15}; \citealt{McClintock:15}; \citealt{Buderman:16}).  Incorporating non-Gaussian data models in the FMM framework so that the procedure remains computationally efficient is a high priority for future research.   

Furthermore, because the computational approach we presented is feasible for large data sets, the individual-level models could be nested within a population-level framework to provide inference for groups of individuals simultaneously (e.g., \citealt{Hooten:16}).   Population-level FMMs with interacting individuals are the subject of ongoing research.

Our study system was somewhat devoid of hard boundaries to movement, but future research to formally incorporate physical constraints to the true position process would be useful (e.g., \citealt{Tracey:05}; \citealt{Brost:15}).  Statistically rigorous methods for accommodating constraints in multivariate continuous-time dynamic processes have been proposed in other fields (e.g., \citealt{CangelosiHooten:09}), but are still nascent in statistical approaches for telemetry data.   

\bigskip
\begin{center}
{\large\bf SUPPLEMENTARY MATERIAL}
\end{center}

\begin{description}
\item[Appendix A:] Derivation of reparameterization of convolved multivariate Brownian motion.  
\item[Appendix B:] Specification and visualization of basis functions in $\tilde{\mathbf{H}}$.  
\item[Appendix C:] Sherman-Morrison-Woodbury identity for FMM. 
\item[Appendix D:] RJMCMC algorithm for fitting FMM. 
\item[Appendix E:] Simulation examples involving homogeneous and heterogeneous dynamics. 
\item[Appendix F:] Empirical variograms for posterior predictive residuals of mule deer and mountain lion models. 
\end{description}

\bibliographystyle{./fullnat}
\bibliography{animalmovement.bib}

\pagebreak

\section*{Appendix A}
Using the previously specified definitions for variables and Ito calculus, we see that the process can be rewritten as:  
\begin{align}
  \boldsymbol\eta(t) &= \int_{t_0}^{t_n} \mathbf{H}(t,\tau) \mathbf{b}(\tau) d\tau \label{eq:intstep1}\\
  &= \int_{t_0}^{t_n} \mathbf{H}(t,\tau) \int_{t_0}^\tau d\mathbf{b}(\tilde{\tau}) d\tau \label{eq:intstep2}\\\
  &= \int_{t_0}^{t_n} \int_{t_0}^\tau \mathbf{H}(t,\tau) d\mathbf{b}(\tilde{\tau}) d\tau \label{eq:intstep3}\\
  &= \int_{t_0}^{t_n} \int_{\tilde{\tau}}^{t_n} \mathbf{H}(t,\tau) d\tau d\mathbf{b}(\tilde{\tau}) \label{eq:intstep4}\\
  &= \int_{t_0}^{t_n} \tilde{\mathbf{H}}(t,\tilde{\tau}) d\mathbf{b}(\tilde{\tau}) \label{eq:intstep5}
\end{align}
\noindent where a step-by-step description for the above is as follows:
\begin{enumerate}   
  \item (\ref{eq:intstep1}): Begin with the convolution model. 
  \item (\ref{eq:intstep2}): Write the Brownian term $\mathbf{b}(\tau)$ in its integral form.  
  \item (\ref{eq:intstep3}): Move the function $\mathbf{H}(t,\tau)$ inside both integrals.  Note that: $t_0<\tilde{\tau}<\tau$ and $t_0<\tau<t_n$.  
  \item (\ref{eq:intstep4}): Switch the order of integration, paying careful attention to the limits of integration.  That is, $\tilde{\tau}<\tau < t_n$ and $t_0 <\tilde{\tau}<t_n$.  
  \item (\ref{eq:intstep5}): Define $\tilde{\mathbf{H}}(t,\tilde{\tau})\equiv\int_{\tilde{\tau}}^{t_n} \mathbf{H}(t,\tau) d\tau$.  The expression can now be written as a convolution of $\tilde{\mathbf{H}}(t,\tau)$ and white noise.      
\end{enumerate}   
\pagebreak

\section*{Appendix B}
The functional forms for the integrated basis functions in Figure 2, are: 
\begin{align}
  \text{BM:  }\; &\tilde{h}(t,\tau)=  1-\Phi\left(\frac{t-\tau}{\phi}\right), \; \phi\rightarrow 0 \\
  \text{IBM:  }\; &\tilde{h}(t,\tau)=
  \begin{cases}
    \frac{\tau-t_1-t}{t_n-t_1} &\mbox{ if } t\leq \tau \\
    0 &\mbox{ if } t > \tau 
  \end{cases} \\
  \text{TU:  }\; &\tilde{h}(t,\tau)=
  \begin{cases}
    1 &\mbox{ if } t_1 \leq t < \tau-\phi \\
    -\frac{2(t-\tau)}{\phi} - \frac{(t-\tau)^2}{\phi^2} &\mbox{ if } \tau-\phi \leq t \leq \tau \\
    0 &\mbox{ if } t > \tau 
  \end{cases} \\
  \text{TD:  }\; &\tilde{h}(t,\tau)=
  \begin{cases}
    1 &\mbox{ if } t_1 \leq t < \tau \\
    1-\left(\frac{2(t-\tau)}{\phi} - \frac{(t-\tau)^2}{\phi^2}\right) &\mbox{ if } \tau \leq t \leq \tau+\phi \\
    0 &\mbox{ if } t > \tau+\phi 
  \end{cases} \\
  \text{G:  }\; &\tilde{h}(t,\tau)=  1-\Phi\left(\frac{t-\tau}{\phi}\right) 
\end{align}

These functional forms for a subset of knots in the temporal domain are shown together in Appendix Figure~\ref{fig:bf_plot}.
\begin{figure}[htp]
  \centering
  \includegraphics[height=6.5in, angle=0]{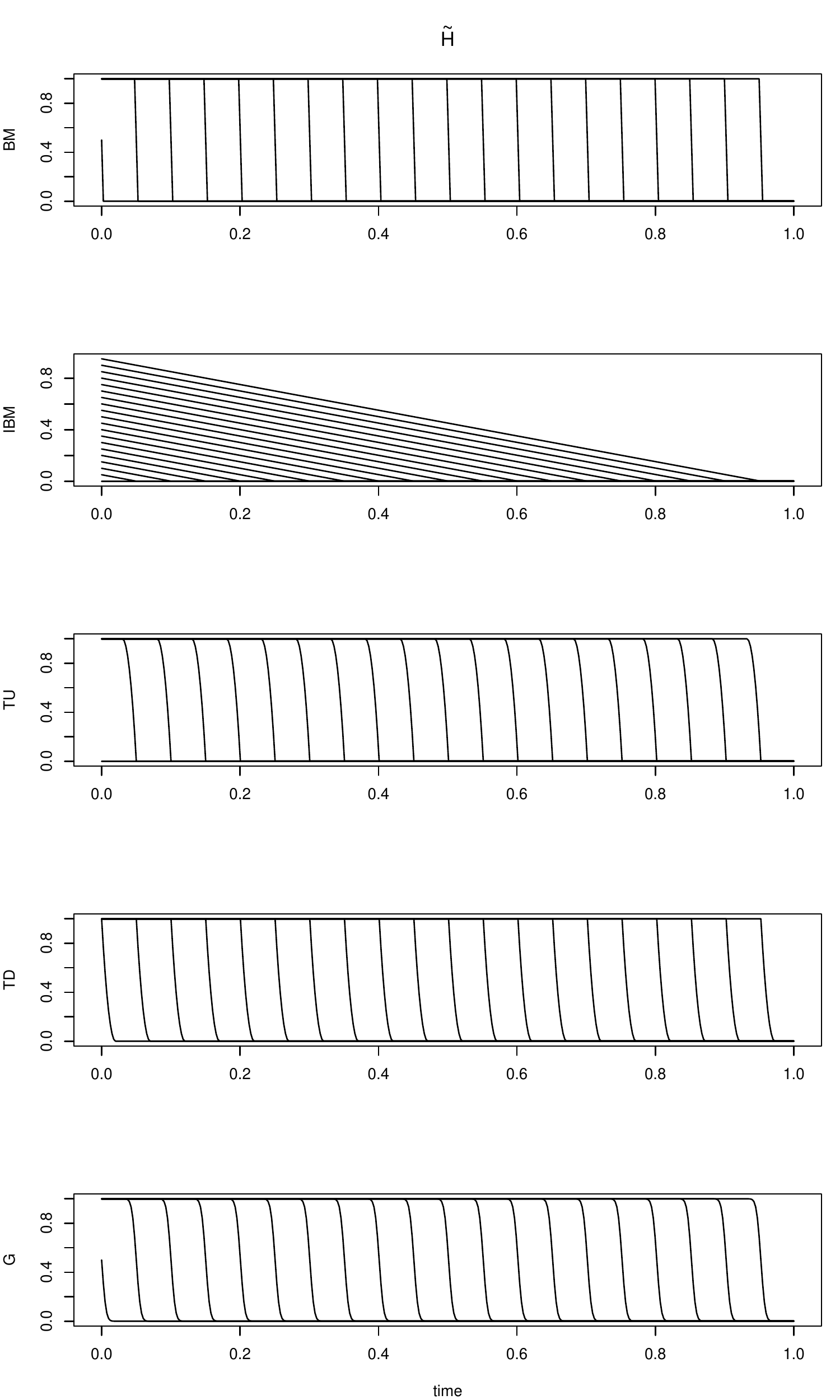}
  \caption{Visualization of 20 basis functions in $\tilde{\mathbf{H}}$.  Individual lines represent rows of $\tilde{\mathbf{H}}$.}
  \label{fig:bf_plot}
\end{figure}
\pagebreak

\section*{Appendix C}
The inverse of the covariance in the integrated FMM can be obtained by
\begin{align}
  (\mathbf{I}\otimes \boldsymbol\Sigma_s + \sigma^2 (\mathbf{I}\otimes \tilde{\mathbf{H}})(\mathbf{I}\otimes \tilde{\mathbf{H}})')^{-1}&= 
  (\mathbf{I}\otimes \boldsymbol\Sigma_s)^{-1} - \sigma^2(\mathbf{I}\otimes \boldsymbol\Sigma_s)^{-1}(\mathbf{I}\otimes \tilde{\mathbf{H}}) \notag \\
  &\times (\mathbf{I}+(\mathbf{I}\otimes \tilde{\mathbf{H}})'(\mathbf{I}\otimes \boldsymbol\Sigma_s)^{-1}(\mathbf{I}\otimes \tilde{\mathbf{H}}))^{-1} \notag \\
  &\times (\mathbf{I}\otimes \tilde{\mathbf{H}})'(\mathbf{I}\otimes \boldsymbol\Sigma_s)^{-1} \notag \; ,
\end{align}
using the Sherman-Morrison-Woodbury identity.  
\pagebreak

\section*{Appendix D}
To implement the RJMCMC approach of \cite{BarkerLink:13}, we use the following algorithm:
\begin{enumerate} 
  \item Set initial MCMC iteration index to $k=1$.
  \item Choose initial model ${\cal M}_j$. 
  \item Select $\boldsymbol\theta \equiv (\phi, \sigma_s^2, \sigma^2)'$ from the $k^{\text{th}}$ iteration of the MCMC output from fitting model ${\cal M}_{lj}$. 
  \item Compute the full-conditional model probability \\
  \begin{equation}
    P({\cal M}_{lj} | \cdot) = \frac{[\mathbf{s}|\boldsymbol\theta_{lj}][\boldsymbol\theta_{lj}|{\cal M}_{lj}]  p_{lj}}{\sum_{{\tilde l}=1}^L \sum_{{\tilde j}=1}^J [\mathbf{s}|\boldsymbol\theta_{\tilde{l}\tilde{j}}][\boldsymbol\theta_{\tilde{l}\tilde{j}}|{\cal M}_{\tilde{l}\tilde{j}}] p_{\tilde{l}\tilde{j}}} \;,  
  \end{equation}
  for each basis function model $l=1,\ldots,L$ and warp $j=1,\ldots,J$.
  \item Sample ${\cal M}_{lj}$ from a categorical distribution with probabilities $P({\cal M}_{11} | \cdot)$, $P({\cal M}_{12} | \cdot)$, $\ldots$ , $P({\cal M}_{LJ} | \cdot)$. 
  \item If inference for the position process is desired, sample $\boldsymbol\mu$ from its full-conditional distribution $[\boldsymbol\mu | \cdot]$, where the full-conditional depends on the current values for the parameters $\boldsymbol\theta$ and the data $\mathbf{s}$.  The full-conditional for $\boldsymbol\mu$ is multivariate Gaussian:
  \begin{equation}
    [\boldsymbol\mu | \cdot] = \mathbf{N}(\boldsymbol\Sigma_{\mu|\cdot}(\mathbf{K}'\boldsymbol\Sigma_s^{-1}\mathbf{s}+\boldsymbol\Sigma_\mu^{-1}(\boldsymbol\mu(0)\otimes \mathbf{1})),\boldsymbol\Sigma_{\mu|\cdot}) \; , 
  \end{equation}
  where,  $\boldsymbol\Sigma_{\mu}\equiv\sigma^2(\mathbf{I}\otimes \tilde{\mathbf{H}})(\mathbf{I}\otimes \tilde{\mathbf{H}})'$ and $\boldsymbol\Sigma_{\mu|\cdot}\equiv(\mathbf{K}'\boldsymbol\Sigma_s^{-1}\mathbf{K}+\boldsymbol\Sigma_\mu^{-1})^{-1}$.
  \item Increment $k=k+1$ and go to step 3.
\end{enumerate}
\pagebreak

\section*{Appendix E}
In this Appendix, we show the results of two examples using simulated data.  In the first example, we simulated a stationary stochastic movement process and, in the second example, we simulated a nonstationary stochastic movement process.  For consistency with the examples involving the mule deer and mountain lion GPS data, we used the same model specification as described in Section 3 in both simulated examples.  

In the first example, we used the Gaussian kernel for $h(t,\tau)$ and the reparameterized covariance matrix 
\begin{displaymath}
  \mathbf{I}\otimes \boldsymbol\Sigma_s + \sigma^2 (\mathbf{I}\otimes \tilde{\mathbf{H}})(\mathbf{I}\otimes \tilde{\mathbf{H}})' = \sigma^2_s (\mathbf{I} + \sigma^2_{\mu/s}(\mathbf{I}\otimes \tilde{\mathbf{H}})(\mathbf{I}\otimes \tilde{\mathbf{H}})'),
\end{displaymath}
\noindent where, $\sigma^2_{\mu/s}=\sigma^2/\sigma^2_s$ and $\tilde{\mathbf{H}}(\phi)$ is a function of the temporal range parameter $\phi$.  For the simulated data, we used a discrete uniform prior for $\phi$ on 100 evenly spaced values from $0.001$ to $0.1$, a uniform prior for $\sigma_{\mu/s}$ on support $(0,20)$, and an informative inverse gamma prior for $\sigma^2_s \sim \text{IG}(12,0.01)$ representing the same strong prior information available for GPS telemetry error similar to that in the mule deer and mountain lion data.  
 
In the first example, we simulated 300 telemetry observations (to mimic the types of data obtained in many telemetry studies) based on the parameter values $\sigma^2_{s}=0.001$, $\sigma^2_{\mu/s}=0.01/\sigma^2_s$, and $\phi=0.005$. The marginal posterior distributions and trace plots for the first example are shown in Appendix Figure~\ref{fig:traceplots}.  The true parameter values were captured well in this simulation example.  We also observed similar inference in other simulations and, for a range of sample sizes, posterior inference was similarly acceptable.  Empirically, we found that the model demonstrated good asympotic properties, providing less biased and more precise inference with larger sample sizes.  This empirical consistency mirrors results from similar continuous process models using basis function constructions in spatial statistics.         
\begin{figure}[htp]
  \centering
  \includegraphics[width=5.5in, angle=0]{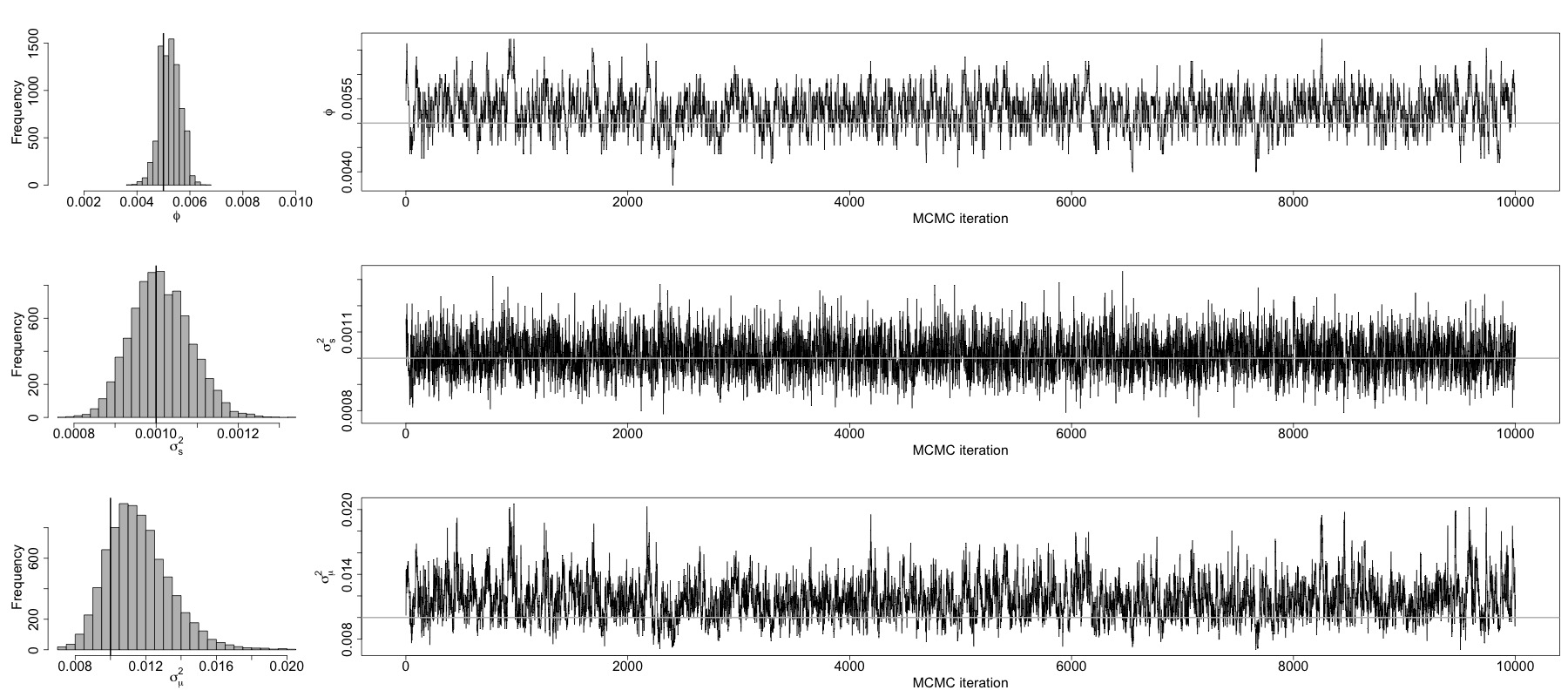}
  \caption{Left panels:  MCMC histograms for model parameters (i.e., approximate marginal distributions); simulation truth shown as vertical black line.  Right panels:  Trace plots exhibiting adequate convergence and mixing.}
  \label{fig:traceplots}
\end{figure}

The mean posterior predictive position process and associated uncertainty are shown in Appendix Figure~\ref{fig:predex1}.  While it is well known that Gaussian process models excel at prediction, the Bayesian version of the model provides a straightforward mechanism for obtaining uncertainty estimates for these complicated continuous-time trajectory processes.
\begin{figure}[htp]
  \centering
  \vspace{-.5in}
  \includegraphics[height=7in, angle=0]{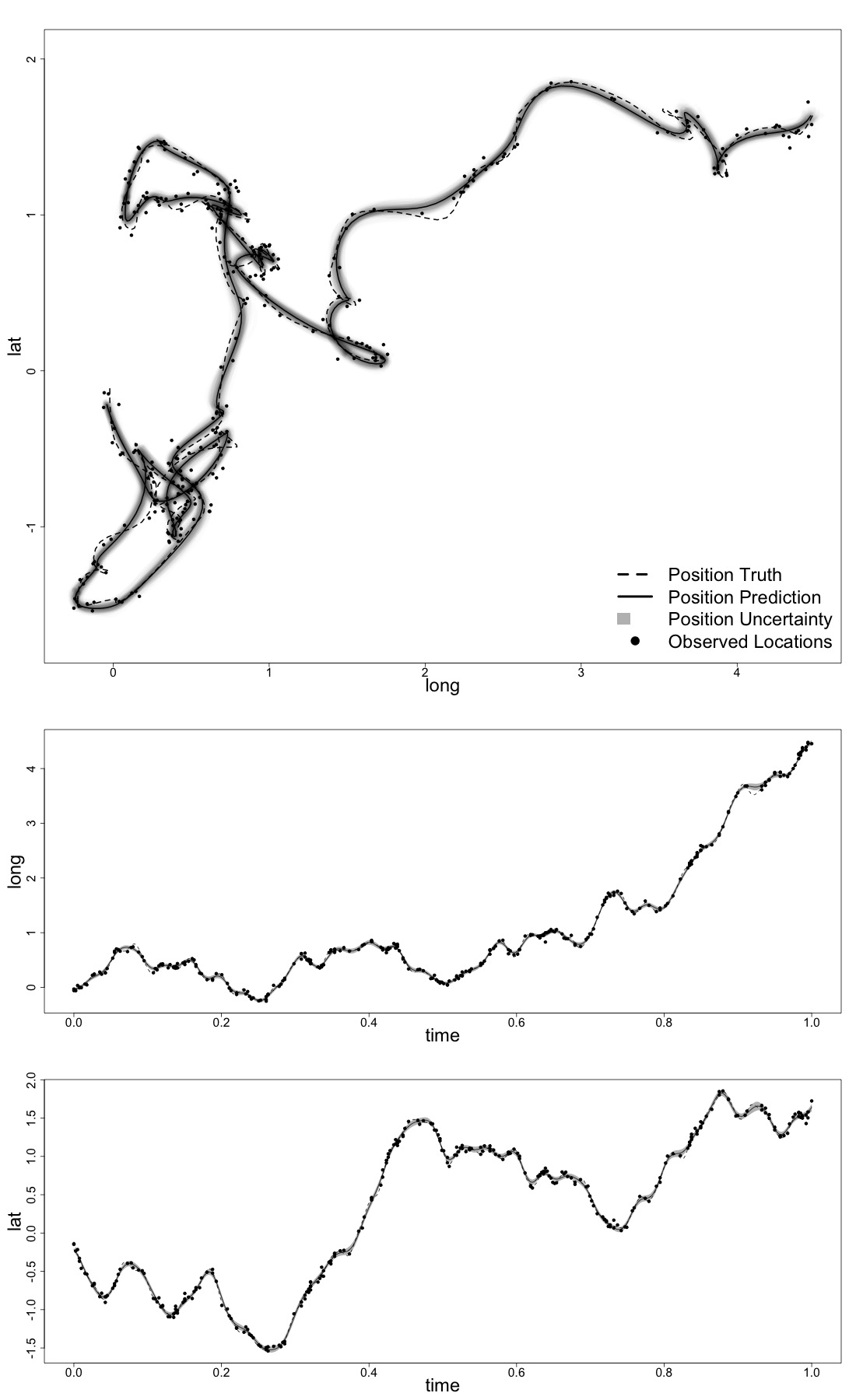}
  \caption{FMM results from data simulated based on a non-warped temporal domain.  Top panel:  Two-dimensional position process and data (predicted in solid, truth in dashed).  Uncertainty associated with the predictive distribution is shown as overlaid predictive realizations of the position process in gray.  Bottom panels:  One-dimensional position process and data (predicted in solid, truth in dashed).  The uncertainty is shown as pointwise 95\% credible intervals for the posterior predictive distribution of the position process.}
  \label{fig:predex1}
\end{figure}
Finally, we checked the model for remaining spatial temporal autocorrelation based on these simulated data.  The empirical posterior predictive residual variogram is provided in Appendix Figure~\ref{fig:homog_vg} and shows no evidence of remaining temporal dependence in the data after accounting for the movement process. 
\begin{figure}[htp]
  \centering
  \includegraphics[height=5in, angle=0]{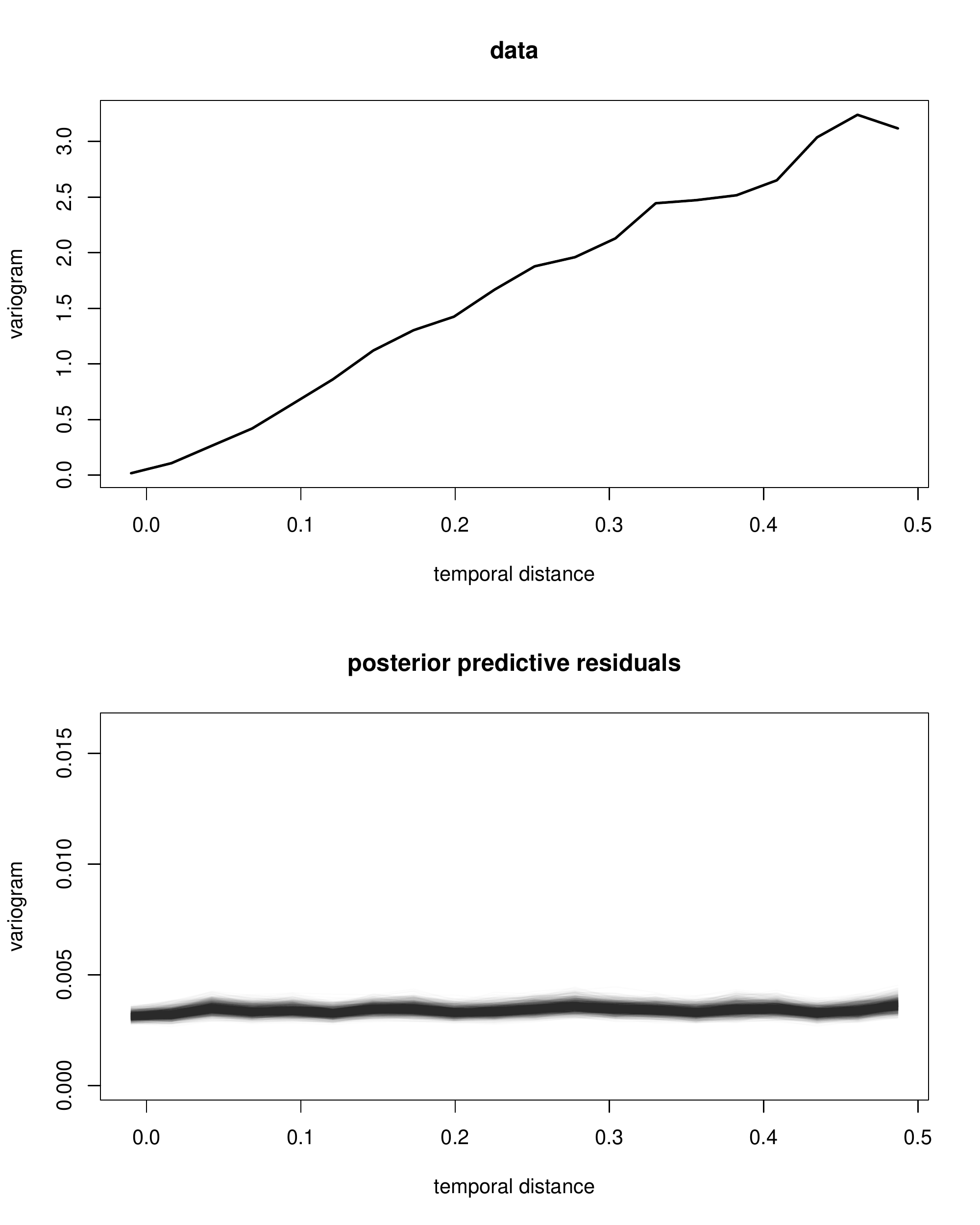}
  \caption{Empirical variograms for the data (top) and the posterior predictive residuals (bottom) of the FMM for the simulated data based on a non-warped temporal domain (posterior predictive realizations indicate uncertainty in the empirical variogram).}
  \label{fig:homog_vg}
\end{figure}

In the second example, we allowed for heterogeneous dynamics in the continuous-time movement process by simulating a temporal warp field $w(t)$ as described in Section 2.4 (shown in Appendix Figure~\ref{fig:predex2}).  We then simulated the continuous-time movement process and data using the same parameters as in the first simulation example described above.  In implementing the FMM, we sampled approximately 4000 warp fields using a Latin hypercube design from a Gaussian process with Gaussian covariance constrained to avoid folding.  The warp distribution was based on 100 parameter combinations for $\sigma_w$ and $\phi_w$ (10 equally spaced values for each parameter, ranging from $0.001-1$).     
\begin{figure}[htp]
  \centering
  \vspace{-.85in}
  \includegraphics[height=7in, angle=0]{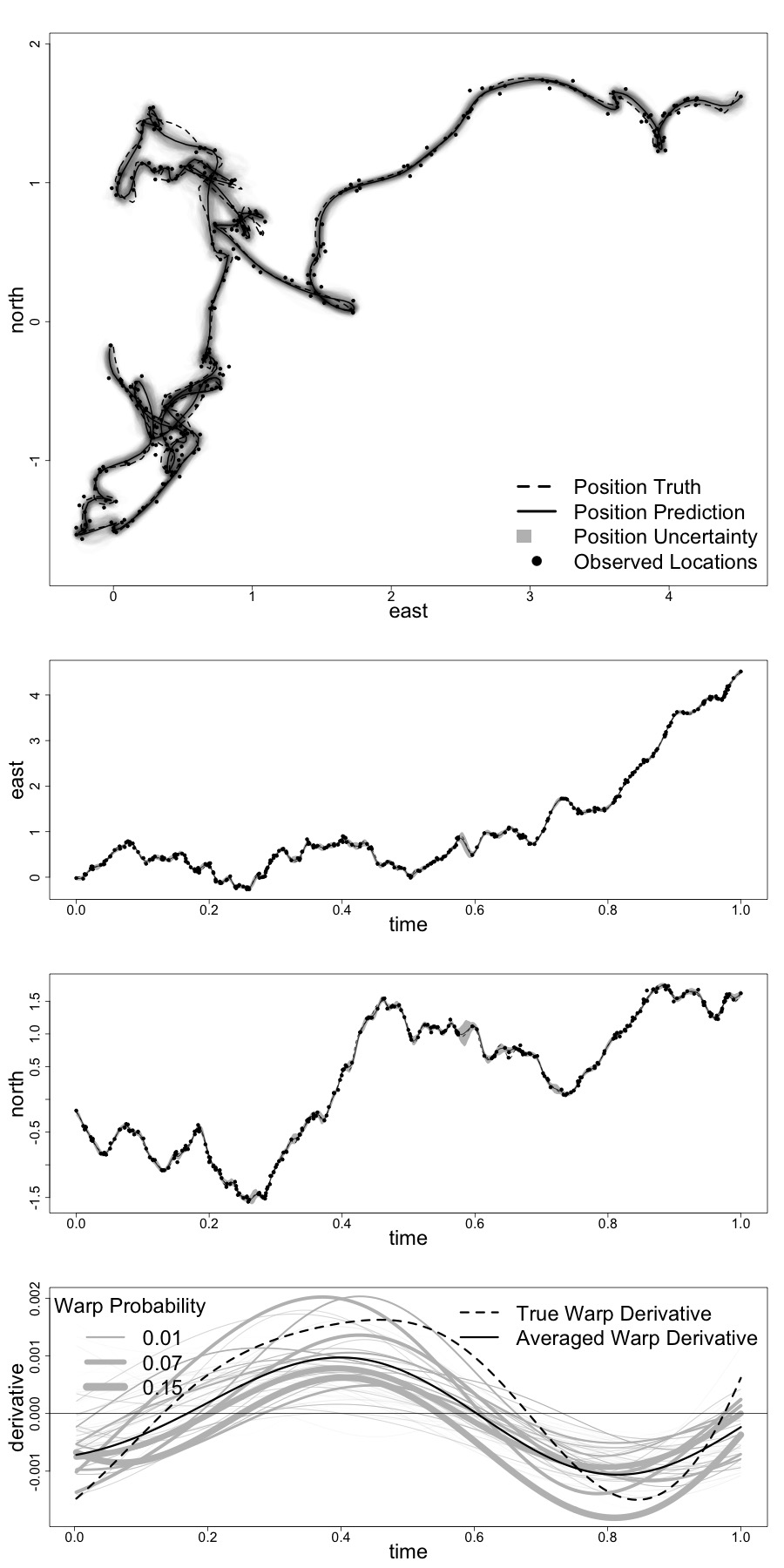}
  \caption{FMM results from data simulated based on nonstationary movement process.  Top panel:  Two-dimensional position process and data (predicted in solid, truth in dashed).  Uncertainty associated with the predictive distribution is shown as overlaid predictive realizations of the position process in gray.  Middle panels:  One-dimensional position process and data (predicted in solid, truth in dashed).  The uncertainty is shown as pointwise 95\% credible intervals for the posterior predictive distribution of the position process.  Bottom panel:  True warp derivative and warp derivative obtained via BMA.  Sample warps shown in gray are weighted by their posterior model probabilities (warps with posterior probability $<0.01$ not shown).}
  \label{fig:predex2}
\end{figure}
As shown in Appendix Figure~\ref{fig:predex2}, the simulated process contains visibly smoother dynamics during the time period 0.7 -- 0.9 and less smooth elsewhere in the temporal domain (except for times before 0.1).  

Using the procedure described in Appendix C of this Supplementary Material, we fit the FMM using the candidate warp fields and all basis functions described in Appendix B and then performed BMA to obtain a final posterior predicted position process and associated warp derivative (Appendix Figure~\ref{fig:predex2}).  The bottom panel of Appendix Figure~\ref{fig:predex2} indicates that the BMA procedure, using approximately 4000 candidate warps, was able to recover the general pattern of nonstationarity in the process (i.e., smoothness before time 0.1 and during the time interval 0.7--0.9).  

\begin{figure}[htp]
  \centering
  \includegraphics[height=5in, angle=0]{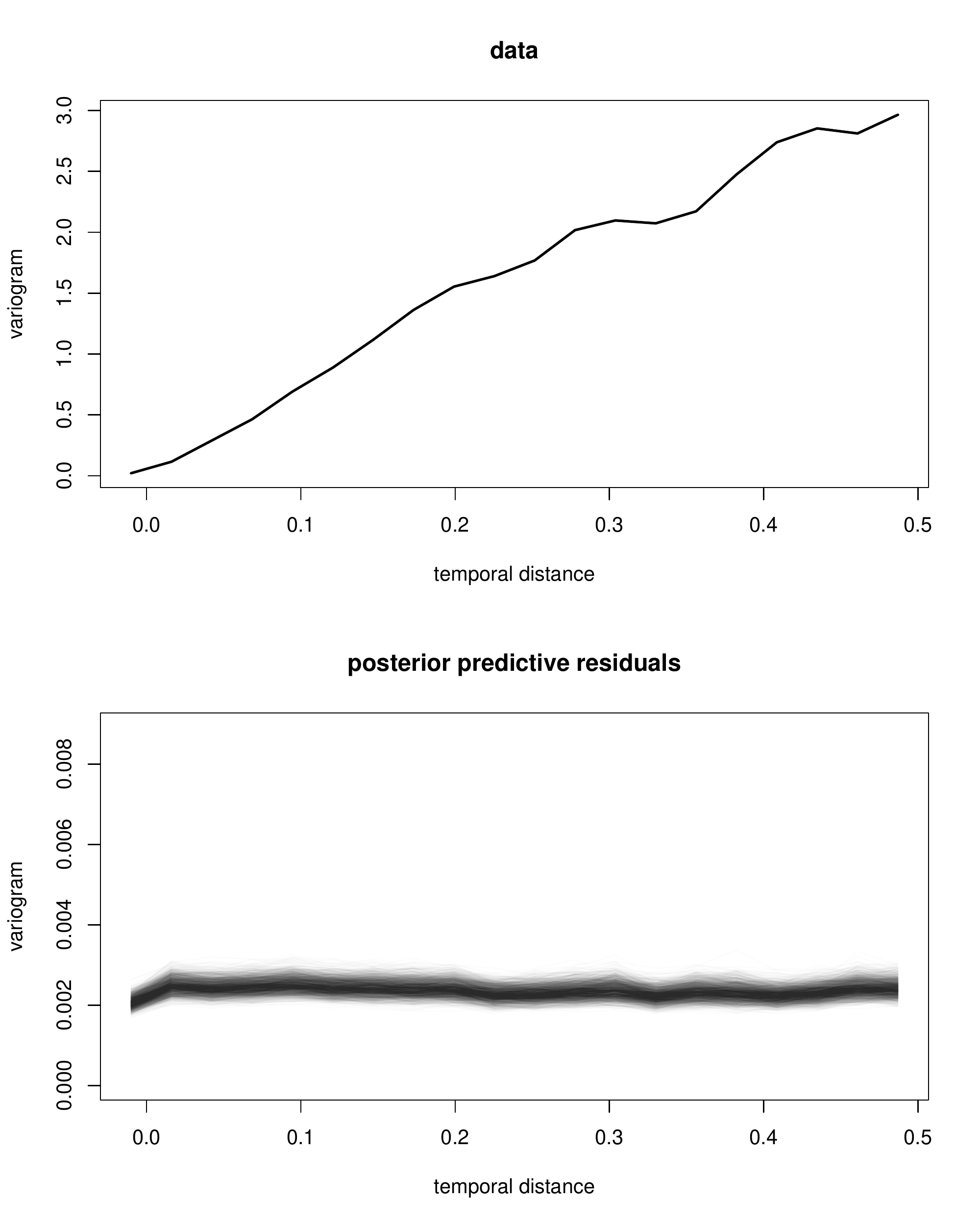}
  \caption{Empirical variograms for the data (top) and the posterior predictive residuals (bottom) of the FMM for the simulated nonstationary data (posterior predictive realizations indicate uncertainty in the empirical variogram).}
  \label{fig:warp_vg}
\end{figure}

The posterior predicted path distribution resulting from the BMA indicates that the FMM recovers the important characteristics of the true position process, despite the temporal gaps in the data.    A variogram for the posterior predictive residuals showed no evidence of remaining temporal structure after fitting the FMM (Appendix Figure~\ref{fig:warp_vg}); that is, the variogram does not contain a significant reduction at short temporal distances.  We observed similar performance over a range of other simulation scenarios as well.       

Finally, the accumulated posterior model probabilities for each basis function (across warps) revealed that models based on Gaussian basis functions accounted for 60\% of the posterior model probability, with tail-up and tail-down basis function FMMs accounting for 39\% and 1\%, respectively.  Brownian and integrated Brownian basis function models accounted for zero posterior model probability in our simulated data example.  

\section*{Appendix F}
The empirical variograms for the posterior predictive residuals based on FMM fits to the mule deer and mountain lion data are shown in Appendix Figures~\ref{fig:MD_warp_vg} and \ref{fig:ML_warp_vg}. 
\begin{figure}[htp]
  \centering
  \includegraphics[height=5in, angle=0]{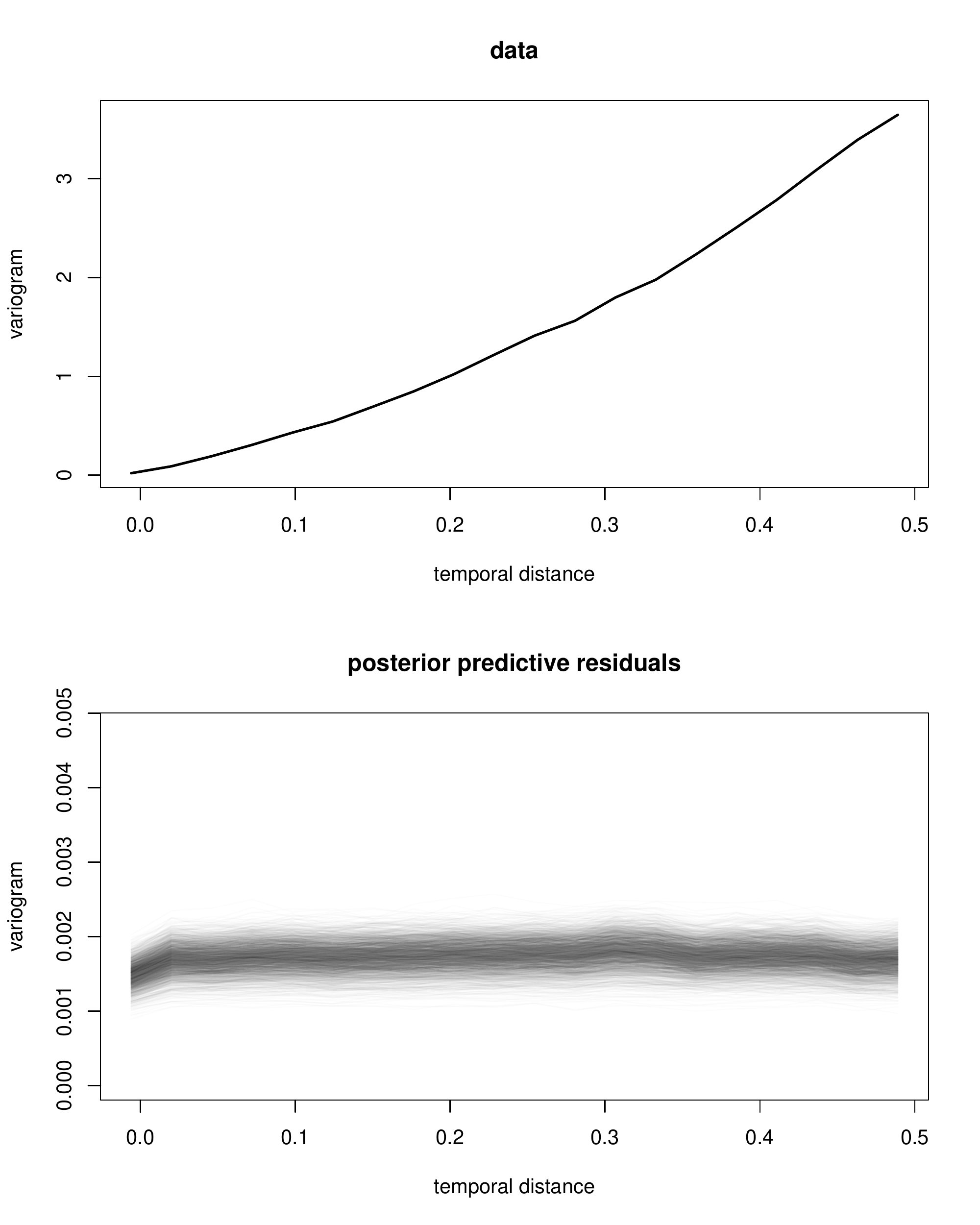}
  \caption{Empirical variograms for the data (top) and the posterior predictive residuals (bottom) of the FMM for the mule deer telemetry data (posterior predictive realizations indicate uncertainty in the empirical variogram).}
  \label{fig:MD_warp_vg}
\end{figure}

\begin{figure}[htp]
  \centering
  \includegraphics[height=5in, angle=0]{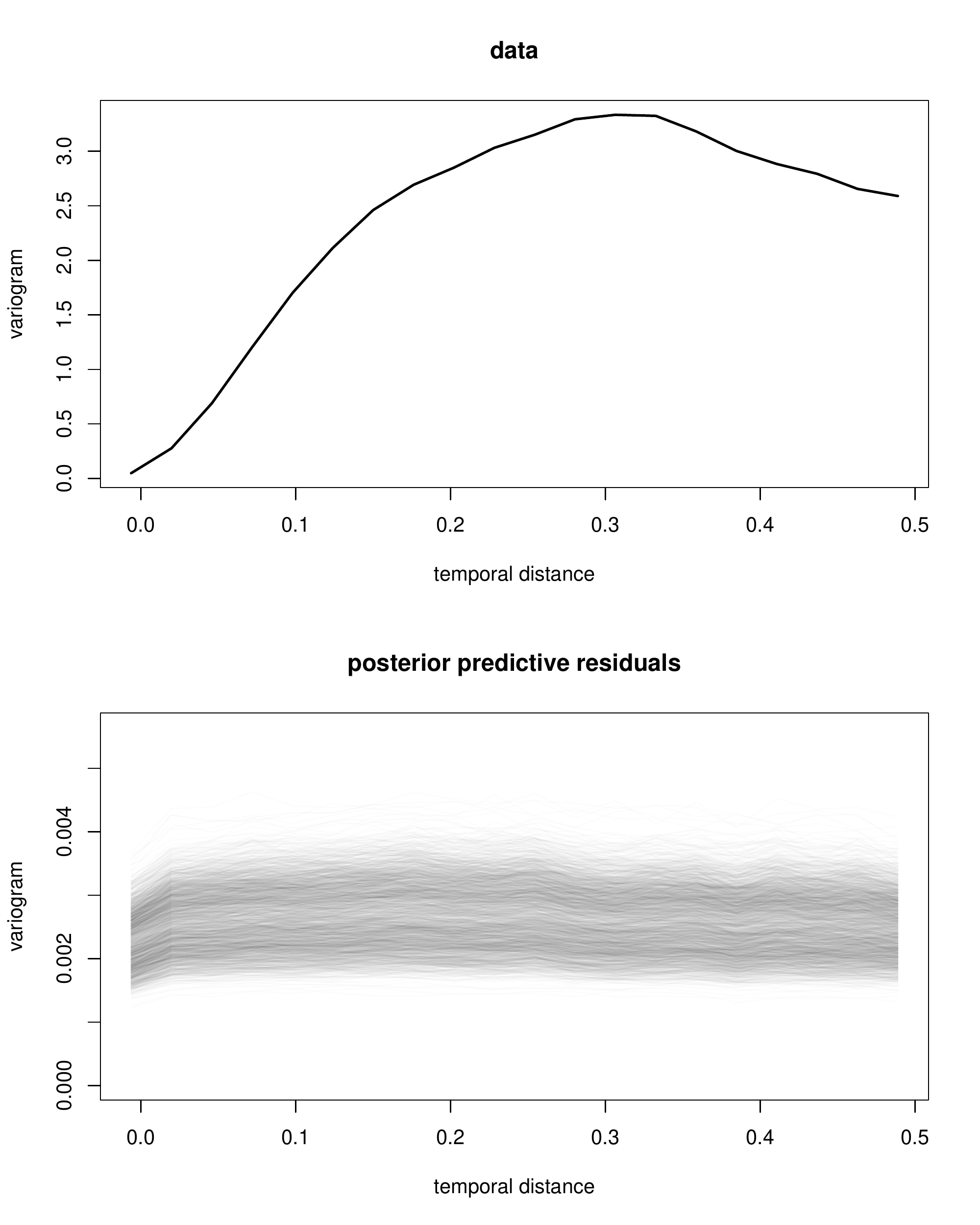}
  \caption{Empirical variograms for the data (top) and the posterior predictive residuals (bottom) of the FMM for the mountain lion telemetry data (posterior predictive realizations indicate uncertainty in the empirical variogram).}
  \label{fig:ML_warp_vg}
\end{figure}

\end{document}